\documentclass[11pt,aps,prf,final,superscriptaddress,longbibliography]{revtex4-1}
\usepackage{hyperref}
\hypersetup{
    colorlinks,
    linkcolor={red},
    citecolor={blue},
    urlcolor={blue}
}
\usepackage{rotating}
\usepackage{mwe}
\usepackage{graphicx}
\usepackage{amsmath,amssymb}
\usepackage{epstopdf, epsfig}
\usepackage{fancyhdr}
\usepackage{graphicx}
\usepackage{caption}
\usepackage{subcaption}
\usepackage{mathrsfs}
\usepackage{enumerate}
\usepackage{wasysym}
\usepackage{multirow}
\usepackage{soul}
\usepackage{natbib}

\usepackage[usenames, dvipsnames]{color}
\definecolor{r}{rgb}{1, 0, 0}
\definecolor{orange}{rgb}{1, 0.5, 0}
\definecolor{vert}{rgb}{0, 0.5, 0}
\definecolor{adb}{rgb}{0, 0.5, 0.5  }

\begin{document}


\title{Spinning and tumbling of long fibers in isotropic turbulence}



\author{Theresa B. Oehmke}
\affiliation{Department of Civil and Environmental Engineering, University of California Berkeley, USA}

\author{Ankur D. Bordoloi}
\affiliation{Department of Civil and Environmental Engineering, University of California Berkeley, USA}
\altaffiliation{Institut des sciences de la Terre, University of Lausanne, Lausanne 1015, Switzerland}

\author{Evan Variano}
\affiliation{Department of Civil and Environmental Engineering, University of California Berkeley, USA}

\author{Gautier Verhille}
\email[]{gautier.verhille@irphe.univ-mrs.fr}
\affiliation{Aix Marseille Univ, CNRS, Centrale Marseille, IRPHE, F-13013 Marseille, France}


\date{\today}

\begin{abstract}
We simultaneously measure both spinning and tumbling components of rotation for long near-neutrally buoyant fibers in homogeneous and isotropic turbulence. The lengths and diameters of the measured fibers extend to several orders of the Kolmogorov length of the surrounding turbulent flow. Our measurements show that the variance of the spinning rate follows a -4/3 power law scaling with the fiber diameter ($d$) and is always larger than the variance of the tumbling rate. This behavior surprisingly resembles that observed previously for sub-Kolmogorov fibers. The general picture that emerges from this study is that long fibers preferentially align with vortex filaments that can be as long as the integral length of turbulence. We compute the Lagrangian time scale and the distribution of both tumbling and spinning that supports this outlook. Our measurements also allow us to quantify the importance of the Coriolis term on the rotational dynamics of fibers in turbulent flows.
\end{abstract}


\maketitle

\section{Introduction}
\label{sec01}
Since 2010, an increasing number of studies are being devoted to the understanding of rotation of anisotropic particles in turbulent flows. The growing interest in this research can be attributed to the numerous applications of such particles found in the environment as well as in industries. The tumbling of elongated fibers are important in paper-making processes. Examples of such applications are also found in polymer processing~\citep{Jarecki2012}, fiber-reinforced-composite molding~\citep{Yashiro2012}, turbulent drag reduction strategies~\citep{MCCOMB1979}, etc.

In real application, most particles are anisotropic ranging from simpler ones, {such as} rods and discs to much more complex shapes~\citep{Voth2017}.  Considering one of the simplest scenarios of  an axisymmetric fiber, {the} rotation can be decomposed into two motions: the tumbling, which corresponds to the rotation of the axis of symmetry of the particle, and the spinning, which corresponds to the rotation about that axis. The evolution of {the} variance of the tumbling rate as a function of fiber length ($L$) has been studied in details in several experimental and numerical works. These studies show that the variance of the tumbling rate for near-neutrally buoyant fibers scales as $\ell^{-4/3}$ when the fiber length is longer than $\sim 10$ Kolmogorov length. The typical lengthscale $\ell$ corresponds to the fiber length $L$ for an aspect ratio $\Lambda=L/d$ larger than $\sim 3$~\citep{Parsa2014, Shin2005}. For smaller aspect ratios $\Lambda \in [1;4]$, \citet{Bordoloi2017} proposed that the pertinent lengthscale is based on the volume of the particle: $\ell\sim(d^2L)^{1/3}$. This scaling was shown to be valid for various shapes with similar aspect ratios~\citep{Pujara2018}. When the fiber inertia cannot be neglected, a filtering effect appears {to decrease} the variance of the tumbling rate~\citep{Bounoua2018, Kuperman2019}. 

Because of the implicit difficulty in resolving both components of rotation, the research heretofore is mainly limited only to the tumbling rate. Using refractive-index-matched PIV, and by analysing the shape of the ellipse produced by the laser sheet intersecting a cylinder, \citet{Bordoloi2017} reported the decomposition of the two components of rotation for cylinders of aspect ratio, $\Lambda = 4$. However, since their experiment was limited only to a single aspect ratio, a complete understanding of the mechanism of rotational partitioning is missing.

The problem also bears an important aspect of fluid mechanics that relates the {rotational} dynamics of anisotropic particles to the velocity gradient tensor in turbulence~\citep{Voth2017}.  Although most studies to date have primarily focused on the dynamics of rigid particles smaller than the Kolmogorov length ($\eta$) ~\citep{Pumir2011,Parsa2012,Chevillard2013,Voth2017},  some have extended this interest to rigid inertial fibers ~\citep{Shin2005,Parsa2014,Bordoloi2017,Bounoua2018,Kuperman2019}, as well as to flexible fibers~\citep{Brouzet2014,Gay2018,Rosti2018,Allende2018,Picardo2020,Sulaiman2019}. Previous theoretical and numerical studies on inertialess fibers shorter than the Kolmogorov length $\eta$ have shown that such small particles  strongly align with the local vorticity~\citep{Pumir2011,Chevillard2013}. As a consequence, small fibers spin more than they tumble~\citep{ Parsa2012, Ni2015, Voth2017}. For fibers longer than the Kolmogorov length $\eta$, such preferential sampling of the velocity field has not been investigated in details. \citet{Pujara2019} showed numerically that when the fiber length $L$ exceeds the viscous regime ($L<\eta$) to the inertial regime ($L>\eta$), the preferential orientation switches from the local vorticity to the most extensional eigen-vector of the coarse-grained strain rate tensor. This could suggest that the spinning rate of a fiber should decrease, as the the preferential orientation with the vorticity is lost when the fiber length is in the inertial regime. {On the contrary, by studying preferential sampling of both flexible and rigid fibers, \citet{Picardo2020} showed that long fibers tend to be preferentially trapped within the vortex tubes in turbulence. In that case, the rate of spinning would increase and might exceed that of tumbling.}

{The goal of this paper is to report direct simultaneous measurements of both spinning and tumbling rates of long inertial fibers in turbulent flows. In the following section \ref{sec02}, we discuss the experimental apparatus and the post-processing methods used to compute the two components of rotation. In section \ref{sec03}, we present the evolution of the tumbling and the spinning rates of these fibers. We analyze and discuss these results in the context of preferential alignment, fiber inertia, time-scale of rotation, and turbulence intermittency in three subsequent subsections. In the final section \ref{sec04}, we conclude with a summary of the key findings of this investigation}.

\section{Experimental setup and methods}
\label{sec02}
\begin{figure}
	\begin{center}
		\centering
	\begin{subfigure}[b]{0.35\textwidth}
		\includegraphics[width=\linewidth]{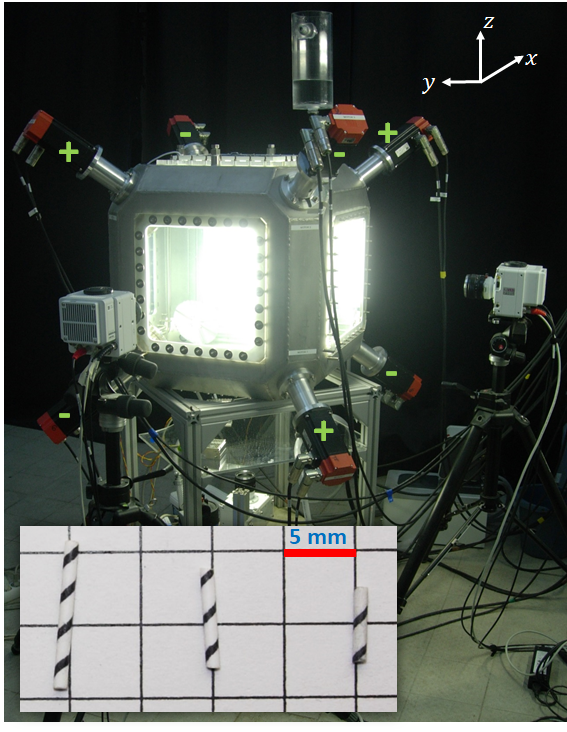}
		\caption{}
		\label{fig:1a}
      \end{subfigure}
	\begin{subfigure}[b]{0.55\textwidth}
		\includegraphics[width=\linewidth]{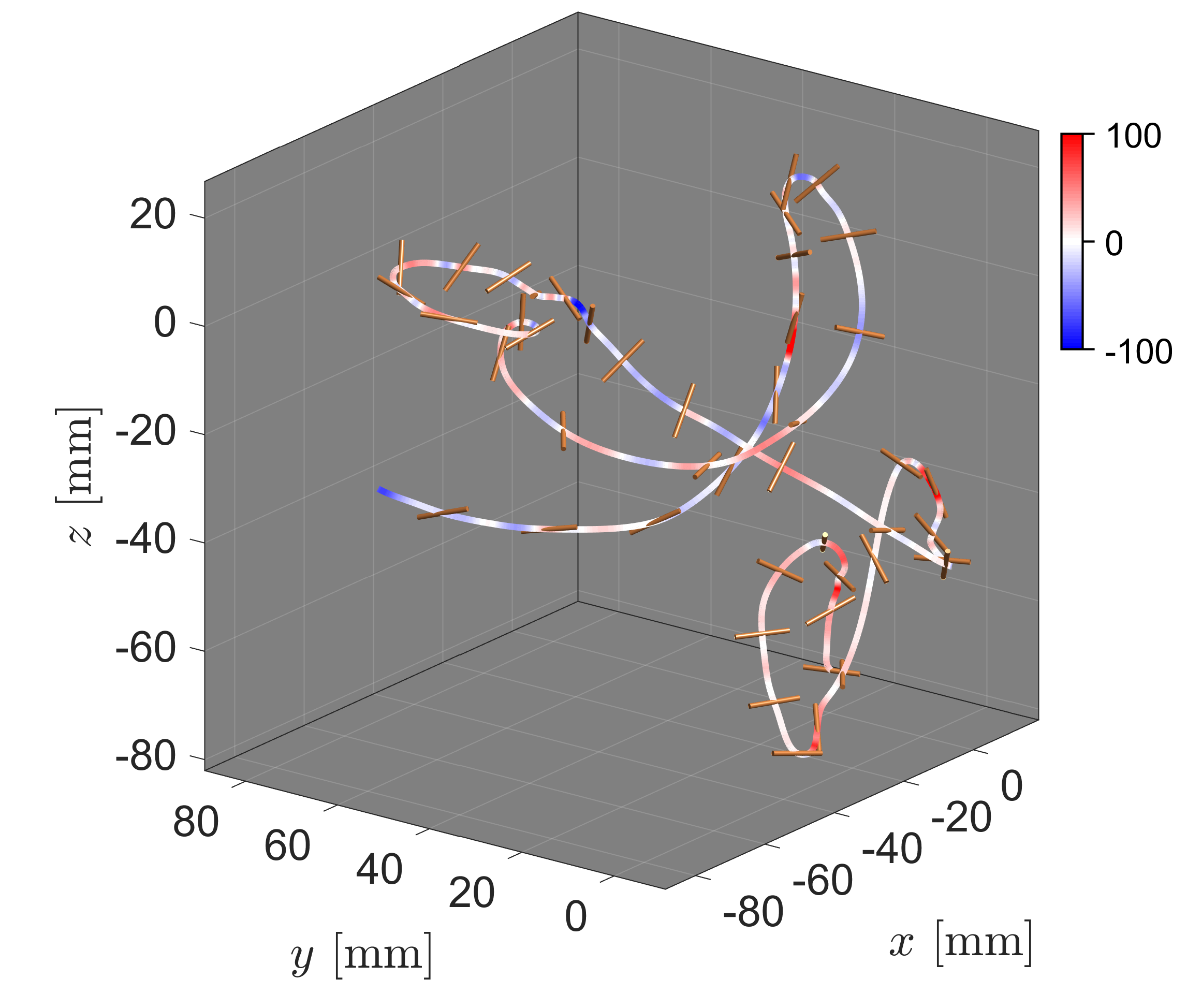}
		\caption{}
		\label{fig:1b}
      \end{subfigure}
		\caption{a) Photograph of the experimental setup. The cubic tank as a side length of 60~cm. Seven of the eight motors are visible. The $+$ or $-$ indicates the direction of the rotation of each impeller. The three cameras are visible along the $x$, $y$ and $z$ axis. The lightning used here is different from the one used for the experiments for artistic reason. On the bottom of the image, the three different kind fibers,10, 7 and 5~mm from left to right, are shown. b) Exemple of 3D trajectory of a 10~mm fibers. The color of the trajectory coded the spinning rate of the fiber (in s$^{-1}$).}
		\label{Fig:ExperimentalSetup}
	\end{center}
\end{figure}

The turbulence is generated by strategically stirring the water filled inside a 60 cm $\times$ 60 cm $\times$ 60 cm  cubic tank. At each corner, an impeller (diameter = 17~cm) with 8 straight blades of thickness 5 mm is driven independently using a 1.5~kW brushless motor. Each impeller is set to rotate at the same frequency but in a chirality opposite to its three nearest neighbors  as shown in FIG.~\ref{Fig:ExperimentalSetup}. The intensity of turbulence inside the tank is set by the impeller frequency between 5 -- 15~Hz. {This configuration allows us to obtain}  homogeneous and isotropic turbulence in a cubic sub-volume of $\approx$ 10~cm $\times$ 10~cm $\times$ 10~cm, centered to the center of the tank. In that zone the mean flow is also negligible (the kinetic energy of the mean flow is around 100 times smaller than the kinetic energy of the turbulent fluctuations). All the measurements presented in this study are performed in this region. Each axis of the reference frame points toward a window and the $z$-axis is parallel and opposite to the direction of gravity, \textit{cf.} FIG.~\ref{Fig:ExperimentalSetup}.

We use two different fluids (pure water and a mixture of water and Ucon) to vary the kinematic viscosity $\nu$ and hence the Kolmogorov length $\eta$ and time $\tau$ on wide ranges. We mix Ucon with water at two concentrations ({approximately 8\% and 11\% by volume }) to increase the liquid viscosity by a factor 6 or 11 from that of water. At the highest concentration used, the fluid density increases  to $\rho_f=1.0147$~kg.m$^{-3}$, which is within 2\% of water density. This change of fluid density is {relatively small and is assumed negligible in the present study}. For all the {tested configurations}, the viscous boundary layer on the impeller is smaller than the height of the blade. Hence, in this range of viscosity, the properties of the turbulence is independent of the kinematic viscosity~\citep{Cadot1998} and can be deduced from the measurement in water using the mechanical properties of the fluid for the mixture of water and Ucon .
Main statistical quantities of turbulence in the volume of measurement are given in the TABLE~\ref{Table:TurbulenceProperties}.
\begin{table}
	\begin{center}
		\begin{tabular}{c |c| c| c| c}
			Integral length & Taylor length & Reynolds number & Kolmogorov length  & Kolmogorov time \\
			$L_I$  & $\lambda$ & $R_\lambda$ & $\eta_K$ & $\tau_K$\\
			$[$cm$]$ & $[$mm$]$ & & $[\mu$m$]$ & [ms]\\
			$ 7$  & $1.7 - 9.7$ & $90-630$ & $34 - 434$ & $1.2-17.1$ \\
		\end{tabular}
		\caption{Turbulence properties for the different experiments presented in that study.}
		\label{Table:TurbulenceProperties}
	\end{center}
\end{table}

We use polystyrene fibers with diameter $d=0.93$~mm and density $\rho_p$=1.04~kg.m$^{-3}$, cut to lengths $L=$5, 7 or 10~mm. Both length and the diameter are in the inertial range of turbulence.  To measure the spinning, a regular helix is printed on the fiber with a pitch of 2.5~mm (see figure~\ref{Fig:ExperimentalSetup}). The tumbling Stokes number, {quantifying the influence of fiber inertia on the tumbling rate,} $St_T=(\rho_p/\rho_f)(d/\eta_K)^{4/3}(d/L)^{2/3}$, defined in~\citet{Bounoua2018}, is always smaller than 2 $\times$10$^{-2}$, such that the inertia of the fiber is negligible at least {for tumbling}. When the carrying fluid is the mixture of water and Ucon, the fiber was coated with a transparent varnish paint (Luxens) to avoid the dissolution of the ink into the fluid. The layer of the paint was thin enough to neglect the modification of the diameter and of the density. In all cases, fibers are slightly heavier than the carrying fluid. However, the settling velocity $U_S\sim (\rho_p-\rho_f)d^2g/16\mu$, where $g$ is the acceleration due to gravity and $\mu$ the dynamic viscosity of the fluid, is at least one order of magnitude smaller than the turbulent fluctuations. Hence, buoyancy effects are negligible here. The volumetric concentration of fibers $\phi$ is small ($\phi<10^{-7}$), so that interaction among fibers and the modification of turbulence by fibers are negligible.

We image the fibers using 3 high speed cameras (Phantom VEO 710L) with  resolution of 1~MPix triggered simultaneously at a frame rate of 1000-3000~fps. The images are captured through a 50~mm lens (Zeiss Planar T 1.4/50) mounted on each camera. The fibers are backlit by an LED panel for the camera pointing along the $z$ axis. Two additional LED spot lights of 6600~lumen are used  to vizualize the pattern printed onto the fiber with the two cameras parallel to the $x$ and $y$ axes. 

	\subsection{Postprocessing}
	
Lagrangian time-series of 3D position and the orientation of each fiber are determined by analyzing the 3 sets of images. To obtain the 3D reconstruction, we determine the translation vector ($\boldsymbol{T}$) and the rotation matrix ($\boldsymbol{R}$) that transform a virtual fiber initially located at the center of the cube ($X_0=\left[0;~0;~0\right]$) {with its} axis of symmetry parallel to the $z$ axis to the location and the orientation of the fiber imaged by each camera. In the fiber frame of reference, the axes of {the} virtual fiber are denoted as $\left[e_1, e_2, e_3\right]$, such that initially these axes coincide with the lab-axes (i.e. $e_1=e_x$, $e_2=e_y$ and $e_3=e_z$). We define a set of points along the virtual helix as $X_h$. 

The cameras are {modeled} with the classical pinhole model. In this model, a camera is characterized by 11  parameters: its position ($X_c$) and the orientation of its frame in the lab frame (determined by three angles of rotation), its focal distances, the coordinates of the projection of the pinhole onto the image, and the skew parameter (for details, see for instance \citep{Verhille2016,Faugeras2001,Hartley2003}). These parameters are determined during a calibration process where a sphere is moved to a known set of locations and imaged by the 3 cameras. The calibration is performed with the fluid inside the cube to take into account the variation of refractive index between the fluid and the air. As the axis of the camera is perpendicular to the viewing window, the distorsions of the ray light due to the refraction at the fluid/plexiglass/air interfaces can be neglected~\citep{Agrawal2012}. We also ensure that the optical distorsions of the lenses are negligible.

Using homogeneous coordinates, as it is classically done in computer vision, the coordinates of a set of points after a rotation by a matrix $\boldsymbol{R}$ and a translation by a vector $T$ is given by:
\begin{equation}
	Q_f= \left(\begin{matrix}
	& & & \\
	& \boldsymbol{R} & & T\\
	& & & \\
	0 & 0 & 0 & 1\end{matrix}\right) Q_0,
\end{equation}
where $Q_0$ and $Q_f$ are the coordinates of a set of points before and after the rotation/translation~\citep{Verhille2016,Faugeras2001,Hartley2003}. We reconstruct each fiber by determining a translation vector $T$ and a rotation matrix $\boldsymbol{R}$ as described below.

The rotation matrix can be decomposed into two terms: $\boldsymbol{R} = \boldsymbol{R}_T \boldsymbol{R}_S$, where $\boldsymbol{R}_S$ is the rotation matrix for spinning (that is, rotation parallel to the $z$ axis), and $\boldsymbol{R}_T$ is the rotation matrix for tumbling. Each matrix is determined independently in two steps. First, we characterize the fiber based on a position vector $T_0$ and an orientation matrix $R_{T,0}$ determined from a "Shape from Silouhette" algorithm, also known as the convex hull volume method~\citep{Zambrano2018}. In this method, a fiber is reconstructed as a set of voxels. $T_0$ is determined from the center of mass of the group of voxels, and $R_{T,0}$ is the rotation matrix which rotates $e_z$ into the vector $n$ connecting the extremities of the group of voxels. In the second step, the position and the orientation of each fiber is optimized through an optimization process similar to~\citet{Bounoua2018}. The cost function to be minimized during the optimization process is the distance between the {projection} of the reconstructed fiber and the ones of the real fiber detected on each image.

The rotation matrix of spinning $\boldsymbol{R}_S$ is determined similarly by minimizing the cost of projection of the virtual helix ($X_h$) onto the two images from the cameras parallel to the $x$ and $y$ axes. To perform this optimization, only the points of the helix really seen by each camera should be considered. These points can be selected knowing {the parameters of the camera and the position} and the orientation of the fiber. At the end of the optimization process, we use the Rodrigues' rotation formula which allows to convert the rotation matrix $\boldsymbol{R}$ into a Rodrigues vector $x_r$ and vice versa. We store the translation vector $T$ and the Rodrigues' rotation vector $x_r$ for further analysis.

\begin{figure}
	\centering
	\begin{subfigure}[b]{0.45\textwidth}
		\includegraphics[width=\linewidth]{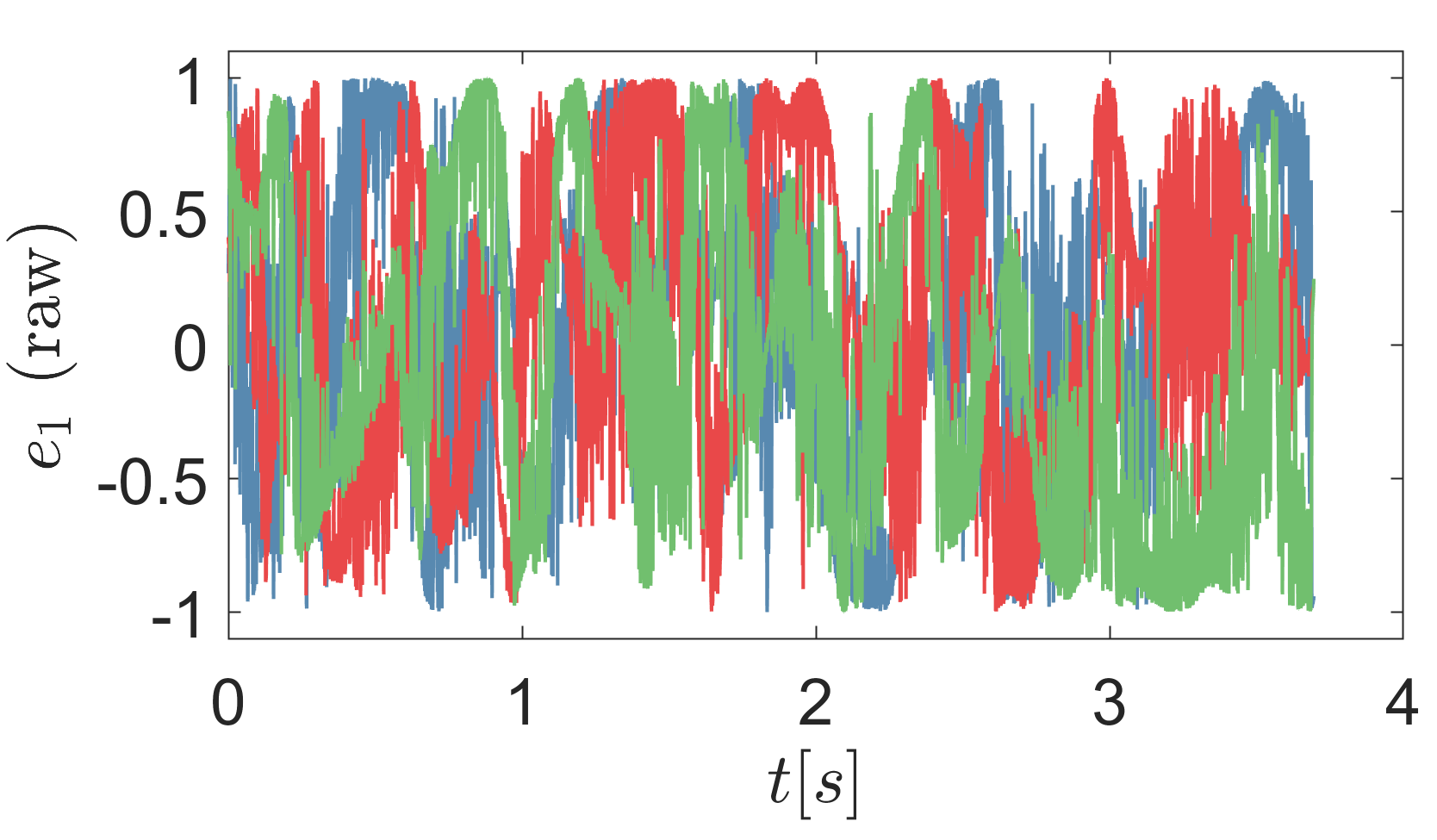}
		\caption{}
		\label{fig:2a}
      \end{subfigure}
	\begin{subfigure}[b]{0.45\textwidth}
		\includegraphics[width=\linewidth]{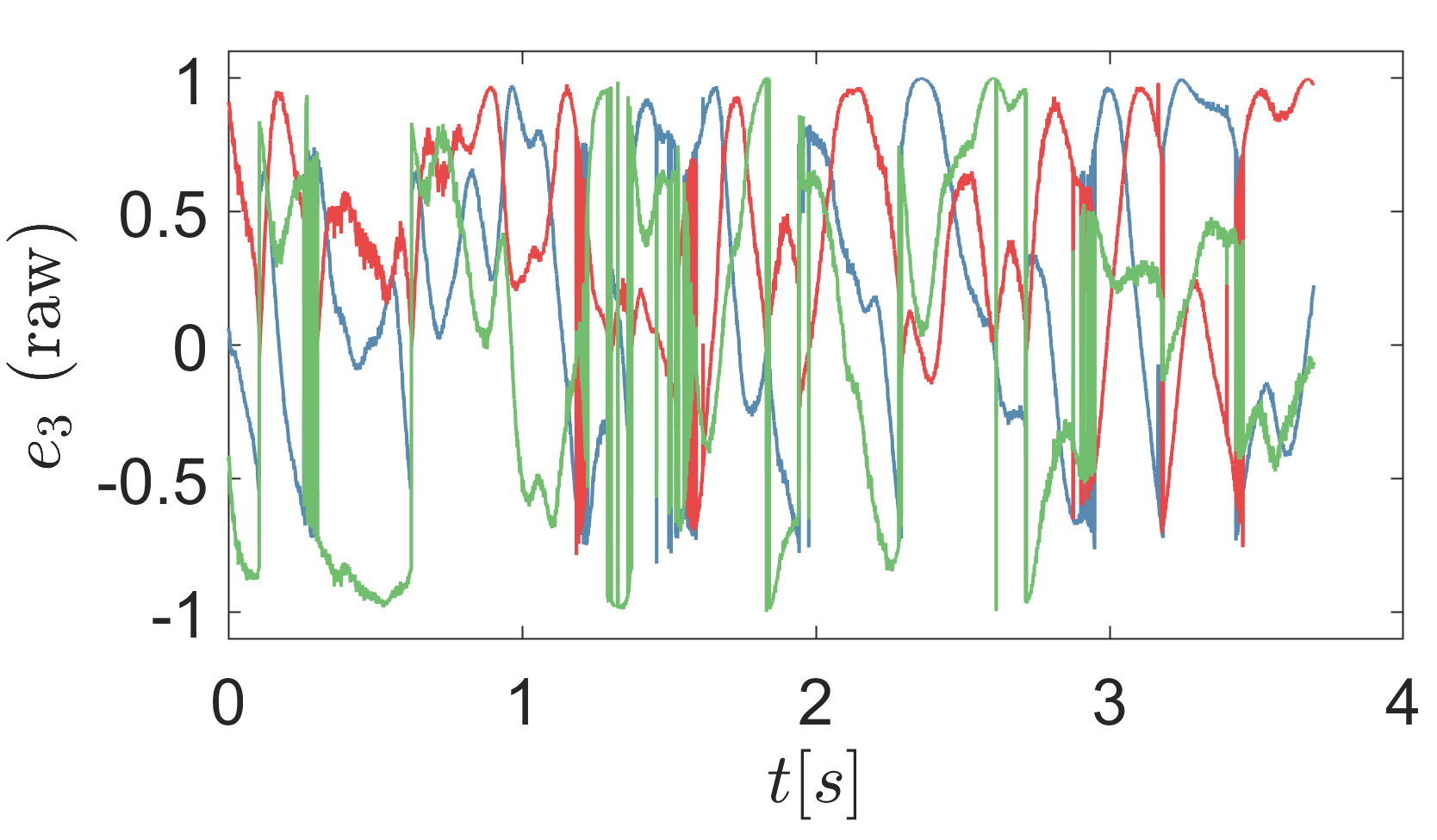}
		\caption{}
		\label{fig:2b}
      \end{subfigure}

	\begin{subfigure}[b]{0.45\textwidth}
		\includegraphics[width=\linewidth]{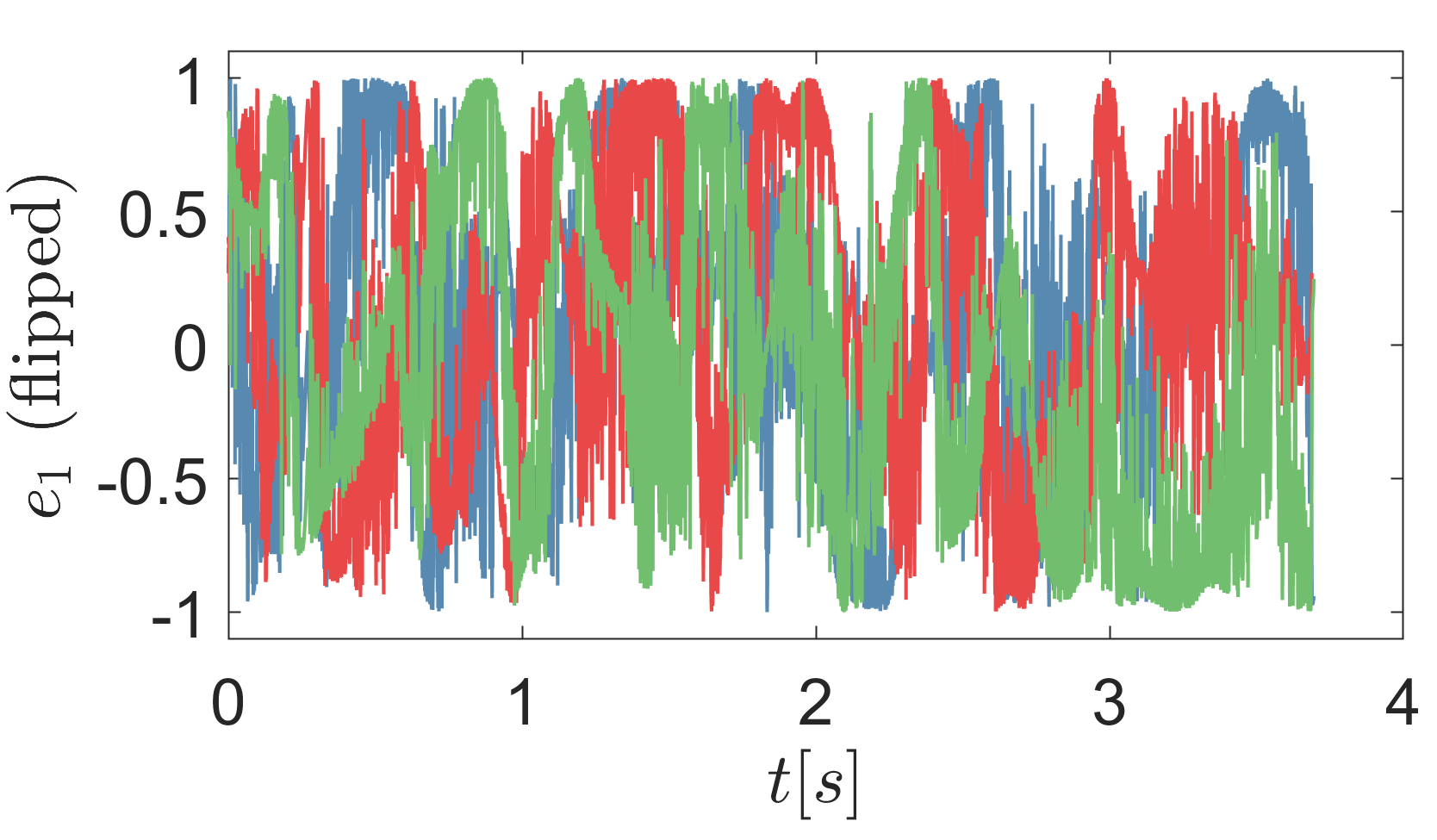}
		\caption{}
		\label{fig:2c}
      \end{subfigure}
	\begin{subfigure}[b]{0.45\textwidth}
		\includegraphics[width=\linewidth]{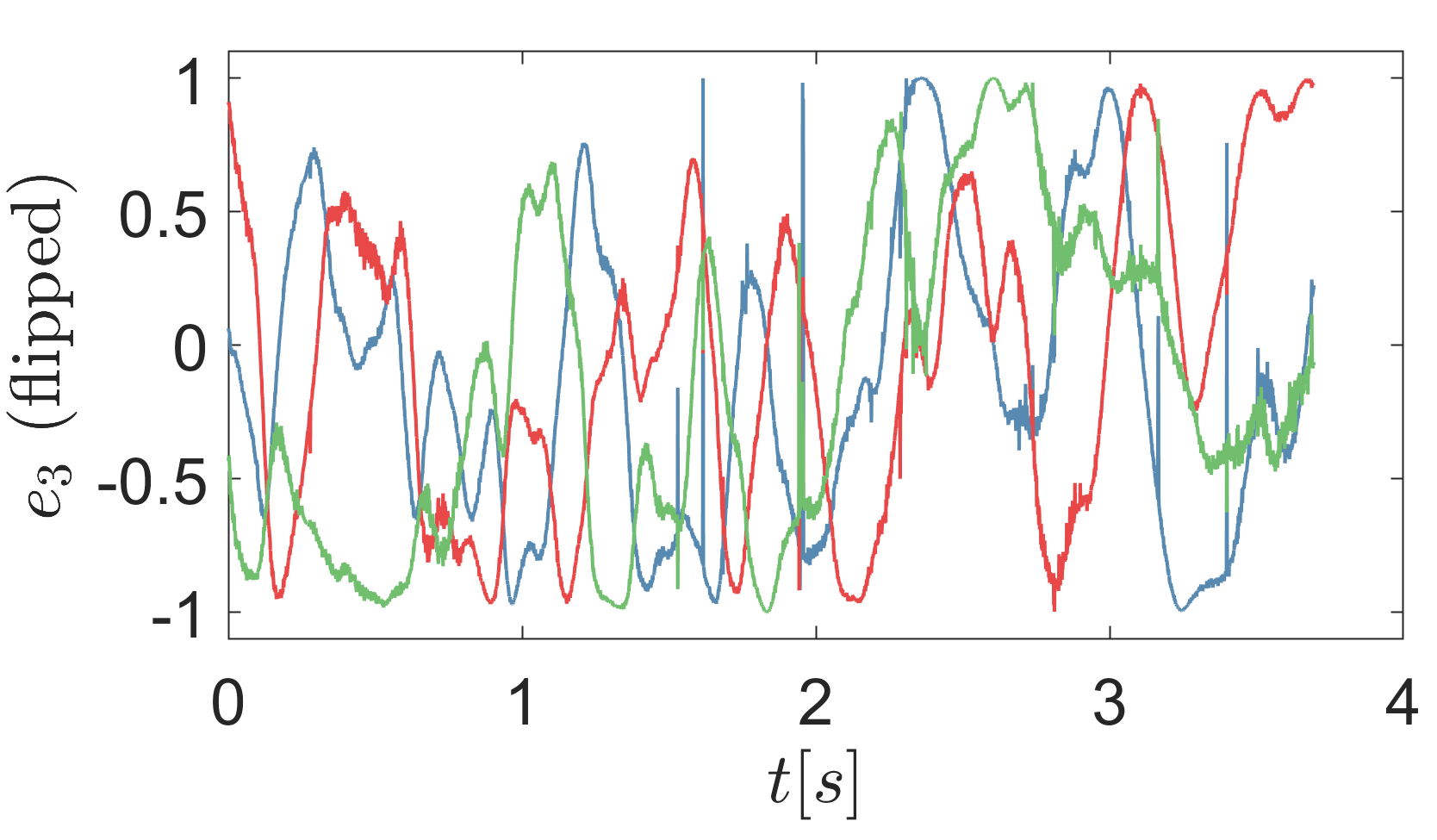}
		\caption{}
		\label{fig:2d}
      \end{subfigure}

	\begin{subfigure}[b]{0.45\textwidth}
		\includegraphics[width=\linewidth]{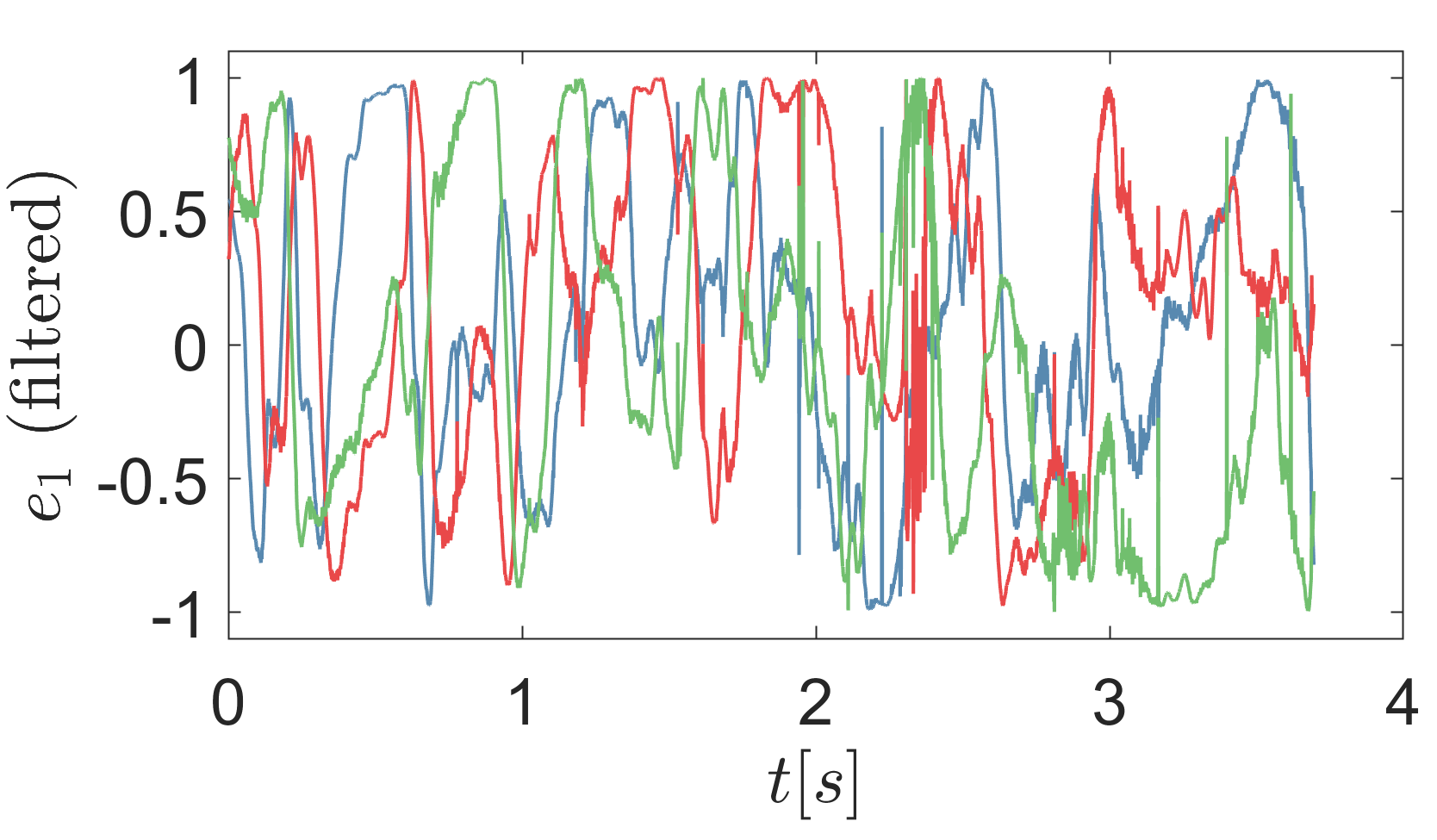}
		\caption{}
		\label{fig:2e}
      \end{subfigure}
	\begin{subfigure}[b]{0.45\textwidth}
		\includegraphics[width=\linewidth]{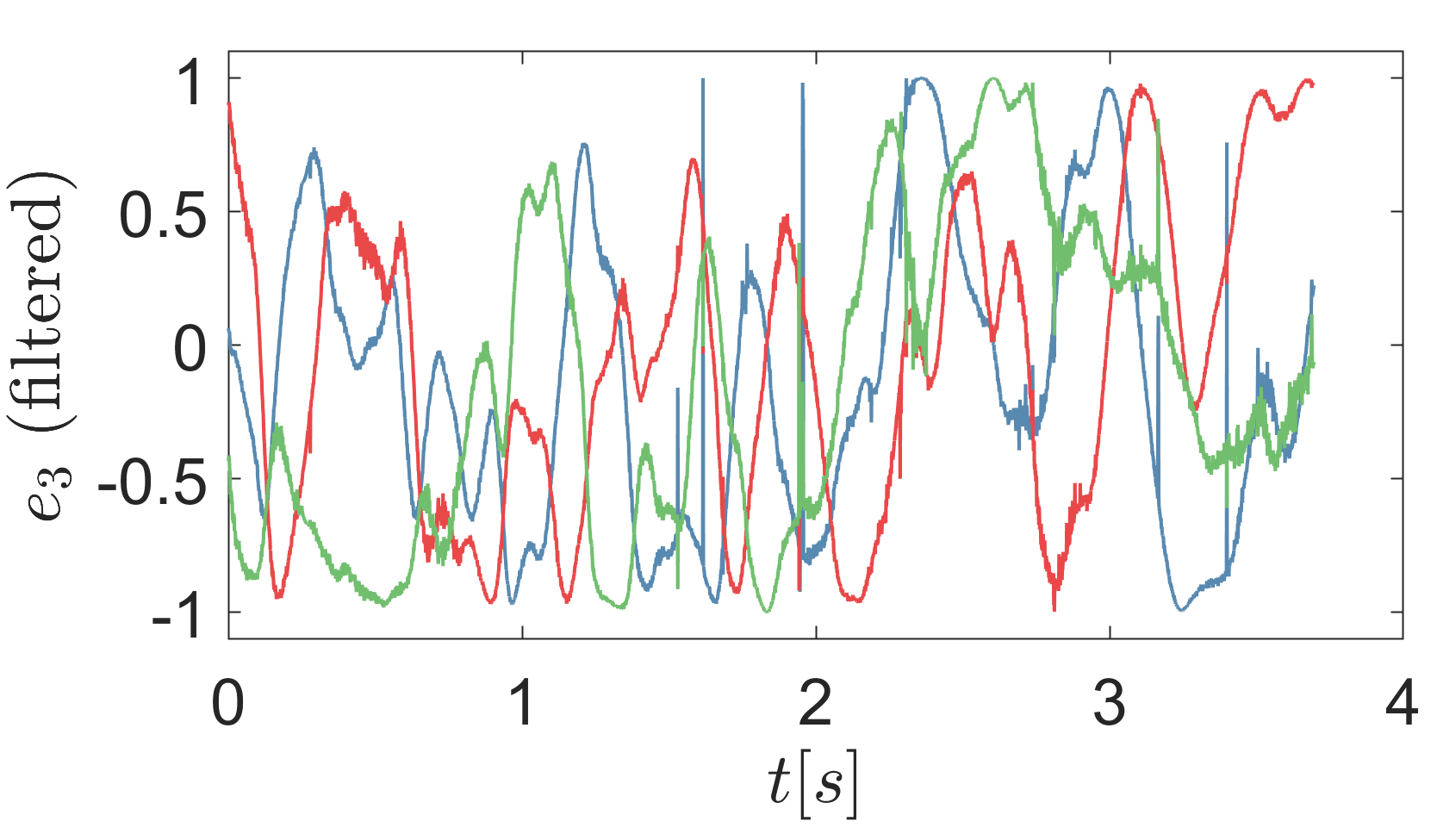}
		\caption{}
		\label{fig:2f}
      \end{subfigure}

		\caption{Time evolution of the three components of $e_1$ (left) and $e_3$ (right) for the raw trajectory (top), after the flipping step (middle) and the smoothing process (bottom). }
		\label{Fig:TrajectoryFiber}

\end{figure}

Once all images are postprocessed, we extract {the} trajectories  of individual fibers using the method of {nearest-neighbor} described in~\citet{Ouellette2005}.   As the concentration of fiber is very low ($< 1\times 10^{-7}$) and the camera acquisition rate is high enough, only the criterion of fiber-fiber distance is used. If several fibers {along the trajectory} satisfy this criterion, the fiber with the orientation closest to that at previous timestamp is selected as a candidate for the trajectory. FIG.~\ref{fig:2a}, \ref{fig:2b}  show the time evolution of the three components of $e_1$ and $e_3$ vectors in the lab frame, respectively for a sample raw trajectory.

The peaks on the trajectory of $e_3$ are due to the ambiguity in the direction (positive vs. negative) of the axis of symmetry ($e_3$) between two successive timestamps. We overcome this ambiguity via a consistency check, such that the direction of $e_3$ of the fiber is flipped if the dot product between $e_3(t)$ and $e_3(t-{\rm d}t)$ is negative (see FIG.~\ref{fig:2c}, \ref{fig:2d}).

As the thickness of the helix is of the order of 1 or 2 pixels, the amplitude of the noise is higher on $e_1(t)$ and $e_2(t)$ than on $e_3(t)$, as seen in the middle panel of FIG.~\ref{Fig:TrajectoryFiber}. A simple approach will be to filter $e_i(t)$ for $i=1,2$, and then to compute the statistics on the filtered data $e_i^f(t)$. This approach however, does not guarantee that the three vectors $e_1$, $e_2$ and $e_3$ form an orthogonal basis. {To overcome this difficulty, we filter the trajectory by applying a Gaussian filter on {$e_1(t)$} and obtaining an optimized $R = R_{opt}$.} The size of the kernel is less than 2$\tau_K$. The amplitude of noise being very small for $e_3(t)$, at each time step we determine an optimal Rodrigues vector that minimizes the distance between $e_1(t)$ and $e_1^f(t)$ with a constraint that $e_3(t)$ remains unchanged. The final $e_1$ and $e_3$ after the smoothing process are shown in  FIG.~\ref{fig:2e}, \ref{fig:2f}.

To obtain convergence in the statistics, only trajectories longer than 100 frames (between 6 and 30 Kolmogorov time depending on the rotation frequency and the fluid viscosity) are used. The longest trajectory is of the order of several seconds for each case, representing several integral time.

\section{Results and discussion}
\label{sec03} 
	\subsection{Tumbling rate}
	\label{sc:3.2}
The tumbling is determined from the variation of the orientation vector $e_3$ using a central difference scheme ($\dot{e}_3$) for each fiber. In the lab reference frame, the tumbling vector $\Omega_T$ can be computed by solving:
\begin{equation}
	\dot{e}_3 = \Omega_T \times e_3 \quad {\rm and} \quad \Omega_T \cdot e_3=0.
	\label{eq:DefTumbVect}
\end{equation}

\begin{figure}
	\begin{center}
		\includegraphics[width=0.5\linewidth]{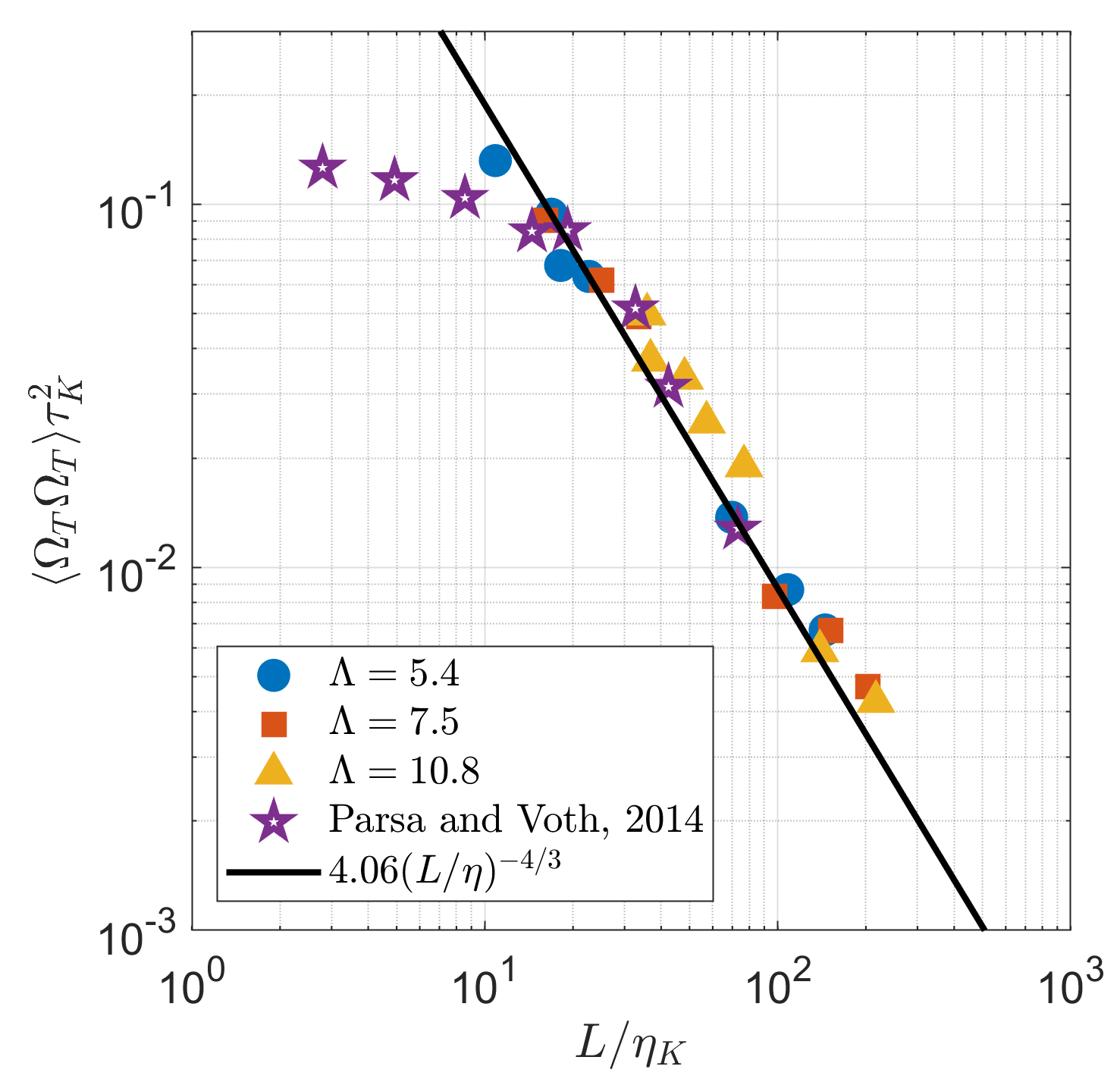}
		\caption{Dimensionless variance of tumbling rate of inertial fibers  against dimensionless fiber length. The prefactor in the -4/3 power-law is obtained from a least-square fit of the experimental data. }
		\label{fig:03}
	\end{center}
\end{figure}

FIG.~\ref{fig:03} presents the normalized variance of the tumbling rate ($\langle\Omega_T\Omega_T\rangle\tau_K^2$) with respect to the normalized fiber length ($L/\eta$). These results are compared with those from \citet{Parsa2014} who studied similar cases. Results from both sets of experiments overlap and show an evolution of the variance of the tumbling rate, as $\langle \Omega_T\Omega_T \rangle \sim (L/\eta)^{-4/3}\tau_K^{-2}$, in agreement with the slender body prediction. In this {model}, fiber inertia is neglected so that the global torque $\Gamma$ applied on the fiber is equal to 0. Considering only the viscous torque~\citep{Bounoua2018} the total torque applied on the fiber can be modeled by:
\begin{equation}
	\Gamma=\int_{-L/2}^{L/2} \mu u_g \times s {\rm d}s = 0,
\label{eq:03}
\end{equation}
where $\mu$ is the dynamical fluid viscosity, $u_g= u_f - v$ is the slipping velocity, with $u_f$ and $v$ denoting the fluid and the fiber velocities respectively. The notation $s$ represents the curvilinear coordinate along the fiber whose origin is at the center of mass. 
As the fiber is rigid, the velocity of the fiber in the frame attached to the fiber can be written $v=s\Omega_T$. Hence, the average tumbling rate from equation \ref{eq:03} scales as:
\begin{equation}
	\Omega_T \sim \frac{1}{L^3} \int_{-L/2}^{L/2} u_f\times s {\rm d}s.
\label{eq:04}
\end{equation}
In the framework of Kolomogorov 1941 (K41) theory, only the structure whose length is comparable to the fiber length contribute to the torque. The fluid velocities ($u_L$) at this scale are constant over the fiber length so the integral in equation~\ref{eq:03} vanishes. This integral also vanishes for velocities at a scale much smaller than the fiber length as they are not correlated along the fiber. Hence, the integral reduces to the slender body scaling, $\langle \Omega_T\Omega_T \rangle \sim (u_L/L)^2 \sim \epsilon^{2/3}L^{-4/3}$.

Following this framework, spinning is forced by the difference of velocity of each part of the fiber about its diameter ($d$). In that case, the term $u_f$ appearing in the expression \ref{eq:04}  should be correlated at scale $d$. Hence, the integral of these contributions along the fiber length should vanish for long aspect ratios, or at least the spinning rate ($\Omega_S$) should be much smaller than the tumbling rate ($\Omega_S$) for long fibers. We report our experimental measurements of the spinning rate and probe this outlook in the following section. 

	\subsection{Spinning rate}

\begin{figure}
	\begin{center}
		\centering
	\begin{subfigure}[b]{0.45\textwidth}
		\includegraphics[width=\linewidth]{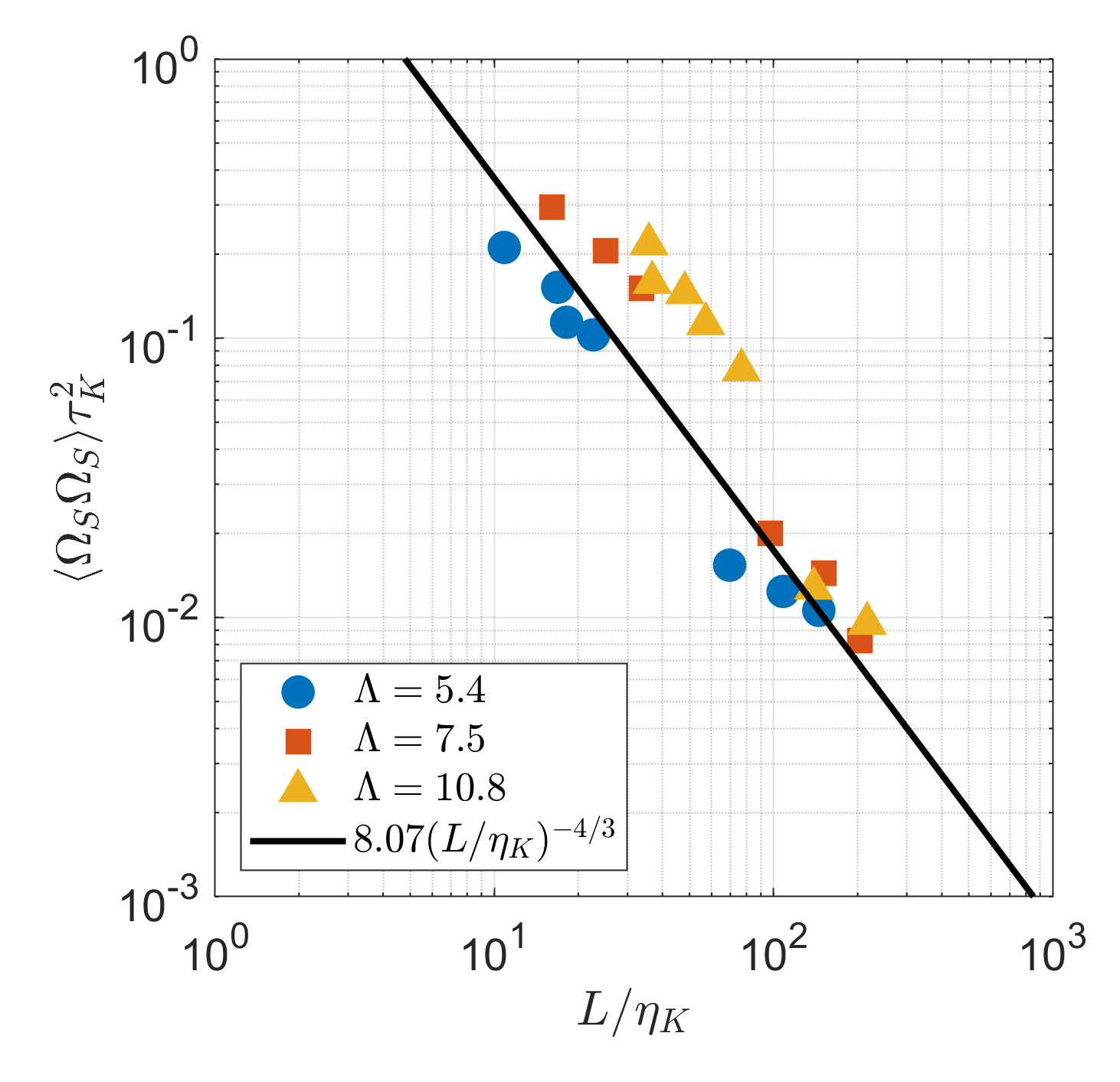}
		\caption{}
		\label{fig:4a}
      \end{subfigure}
	\begin{subfigure}[b]{0.45\textwidth}
		\includegraphics[width=\linewidth]{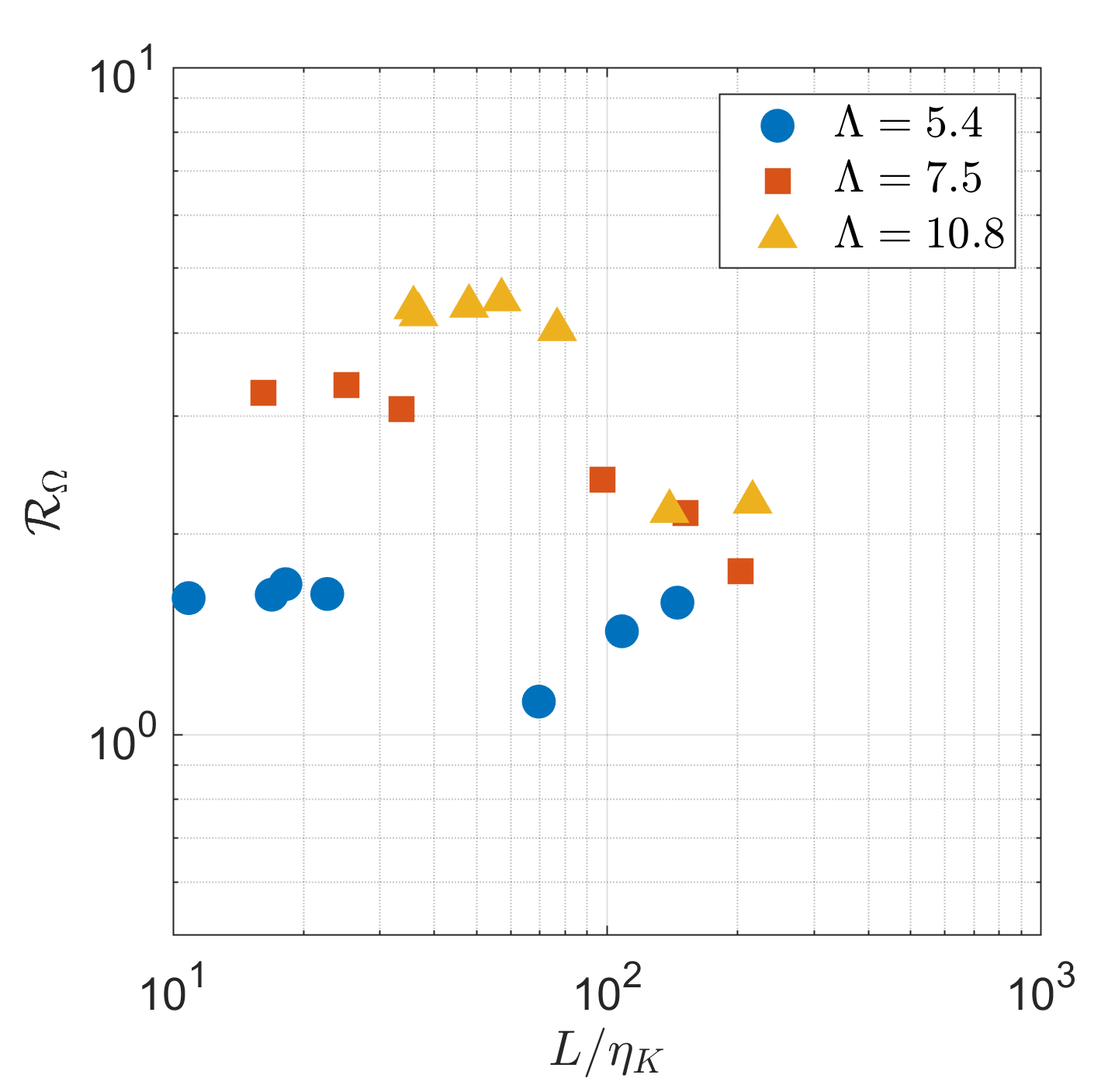}
		\caption{}
		\label{fig:4b}
      \end{subfigure}
		\caption{a) Dimensionless variance of spinning rate and b) the ratio between the variances of spinning and tumbling rates with respect to dimensionless fiber length.}
		\label{Fig:SpinningRate}
	\end{center}
\end{figure}

Contrary to the tumbling rate, the spinning rate cannot be determined directly from the temporal evolution of $e_1$ or $e_2$ as they {depend on both} tumbling and spinning. Therefore, the determination of the spinning rate requires removal of the contribution from tumbling. The rotation matrix $\boldsymbol{R}_T(t)$ related to tumbling is given by the evolution of $e_3(t)$ as:
\begin{equation}
	e_3(t) =\boldsymbol{R}_T(t) e_z,
\end{equation}
enforcing that the rotation axis associated to this matrix is perpendicular to both $e_z$ and $e_3$. Knowing this matrix $\boldsymbol{R}_T$, one can define the spinning rotation matrix $\boldsymbol{R}_S=\boldsymbol{R}_T^{-1}\boldsymbol{R}_{opt}$. The spinning rate ($\Omega_S$) is then determined from the spinning vector $e_S(t)=\boldsymbol{R}_S(t) e_x$ in the {fiber frame} using:
\begin{equation}
	\dot{e}_S=\Omega_S \times e_S.
\end{equation}
Another possibility is to compute directly the total rotation vector $\Omega$ from the temporal evolution of the fiber frame: $\dot{e}_i = \Omega \times e_i$ for $i=1,2,3$. The spinning vector corresponds then to the third component of $\Omega$ and the tumbling to the two first components.  We checked that the presented results are comparable with both method. However, the amplitude of the noise was smaller with the first one.

FIG.~\ref{fig:4a} shows the normalized variance of the spinning rate ($\langle\Omega_S\Omega_S\rangle\tau_K^2$) as a function of the normalized fiber length ($L/\eta_K$). The global trend for the three aspect ratios ($\Lambda$) is a decrease of the {spinning} rate with increasing the normalized fiber length. However compared to $\langle\Omega_T\Omega_T\rangle\tau_K^2$ (see FIG.~\ref{fig:03}), the data points for $\langle\Omega_S\Omega_S\rangle\tau_K^2$ are largely scattered about the -4/3 power-law fit. This suggests that the fiber length is not or not only the controlling parameter of the spinning rate. This is even more evident in the ratio $\mathcal{R}_{\Omega}$ between the variances of spinning and tumbling $\mathcal{R}_\Omega=\langle \Omega_S \Omega_S \rangle / \langle \Omega_T \Omega_T \rangle$ of rotation plotted against the normalized fiber length (see FIG.~\ref{fig:4b}). $\mathcal{R}_{\Omega}$ is roughly constant for each aspect ratio ($\Lambda$), but its value increases with {$\Lambda$}, signifying the effect of the fiber diameter ($d$). FIG.~\ref{fig:4b} also shows that the variance of spinning rate is always larger than that of the tumbling rate. This contradicts the expectation from K41 theory discussed in section \ref{sc:3.2}.

These observations raise two important questions. First, what is the mechanism of forcing behind the spinning of long fibers in turbulence? Second, what is the consequence of the amplitude of the spinning rate on the global rotation dynamics of fibers in turbulence? In earlier studies \citep{Voth2017, Bounoua2018}, the Coriolis term $\Omega\times I\Omega$ was always neglected when modeling the tumbling rate of long fibers. Our current observation urges an investigation related to the validity of this assumption. We address these two questions in the following two sections.

	\subsection{Spinning rate and preferential alignment}
    \label{sc:3C}
\begin{figure}
	\begin{center}
		\centering
		\includegraphics[width=0.5\linewidth]{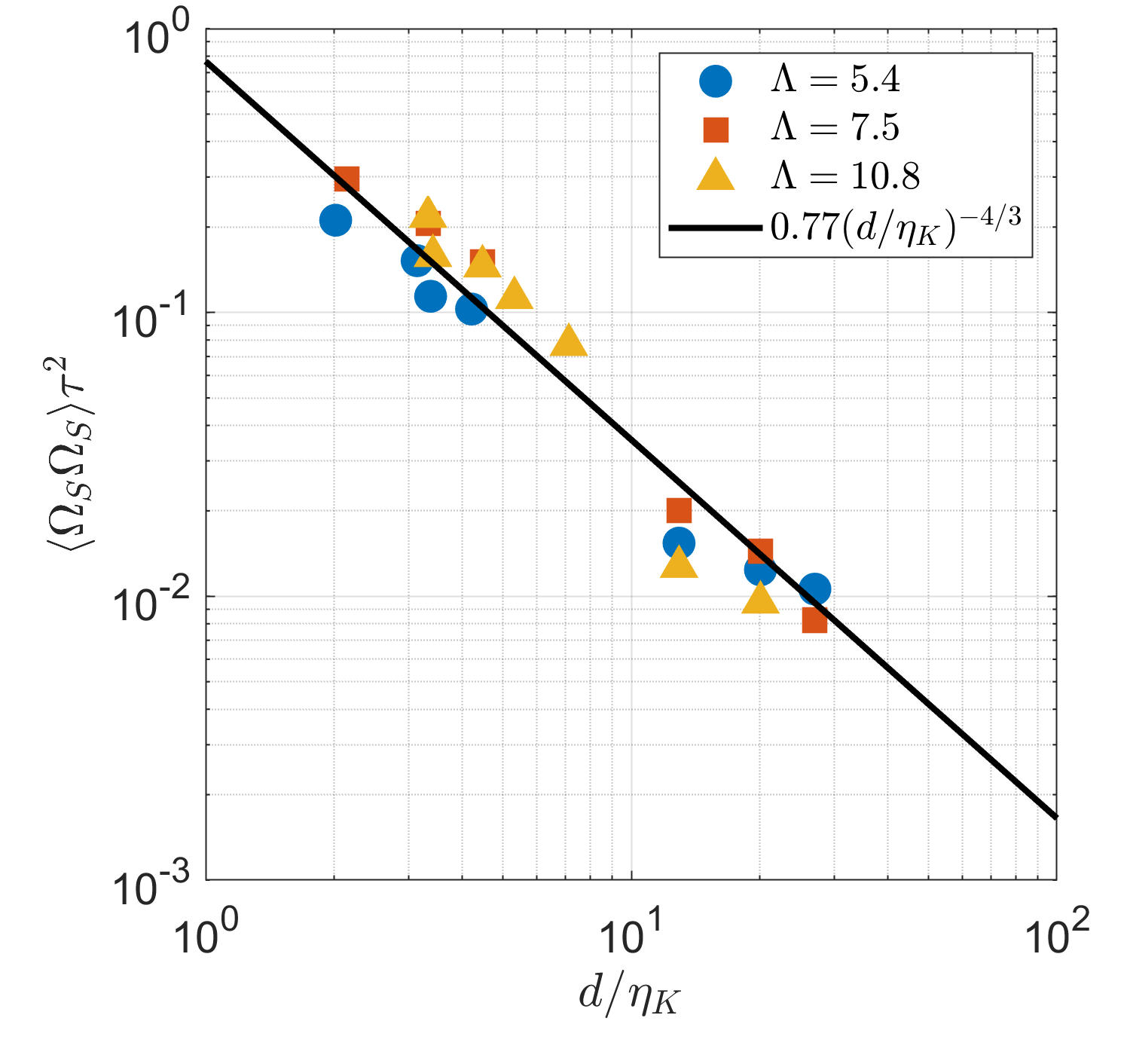}
	\caption{Dimensionless variance  of spinning rate  as a function of the dimensionless fiber diameter.}
\label{fig:5}
	\end{center}
\end{figure}

We examine here two mechanisms that could possibly induce a spinning for a long fiber. In the first one, the spinning is forced by the coarse-grained vorticity. This scenario is incompatible with the results of~\citet{Pujara2019}, where they showed that the alignment of fibers with vorticity decreases with increasing fiber length. Moreover, this scenario requires the spinning rate to scale with the fiber length, which is not compatible with the scattering observed in FIG.~\ref{fig:4a}. The second scenario postulates a preferential alignment with structures that imposes a velocity difference at the scale of the fiber diameter coherent along the fiber length as proposed in \citet{Picardo2020}.  To test this scenario, we show the evolution of the normalized variance of the spinning rate as a function of the normalized fiber diameter in FIG.~\ref{fig:5}. Result indicates that the scattering of the data points is reduced compared to that in FIG.~\ref{fig:4a}. This suggests that the spinning is indeed due to the shear at the scale of the fiber diameter:
\begin{equation}
	\langle \Omega_S\Omega_S \rangle \sim (u_d/d)^{2} \sim (d/\eta)^{-4/3}\tau_{K}^{-2}.
\end{equation}

\begin{figure}
	\begin{center}
		\centering
	\begin{subfigure}[b]{0.45\textwidth}
		\includegraphics[width=\linewidth]{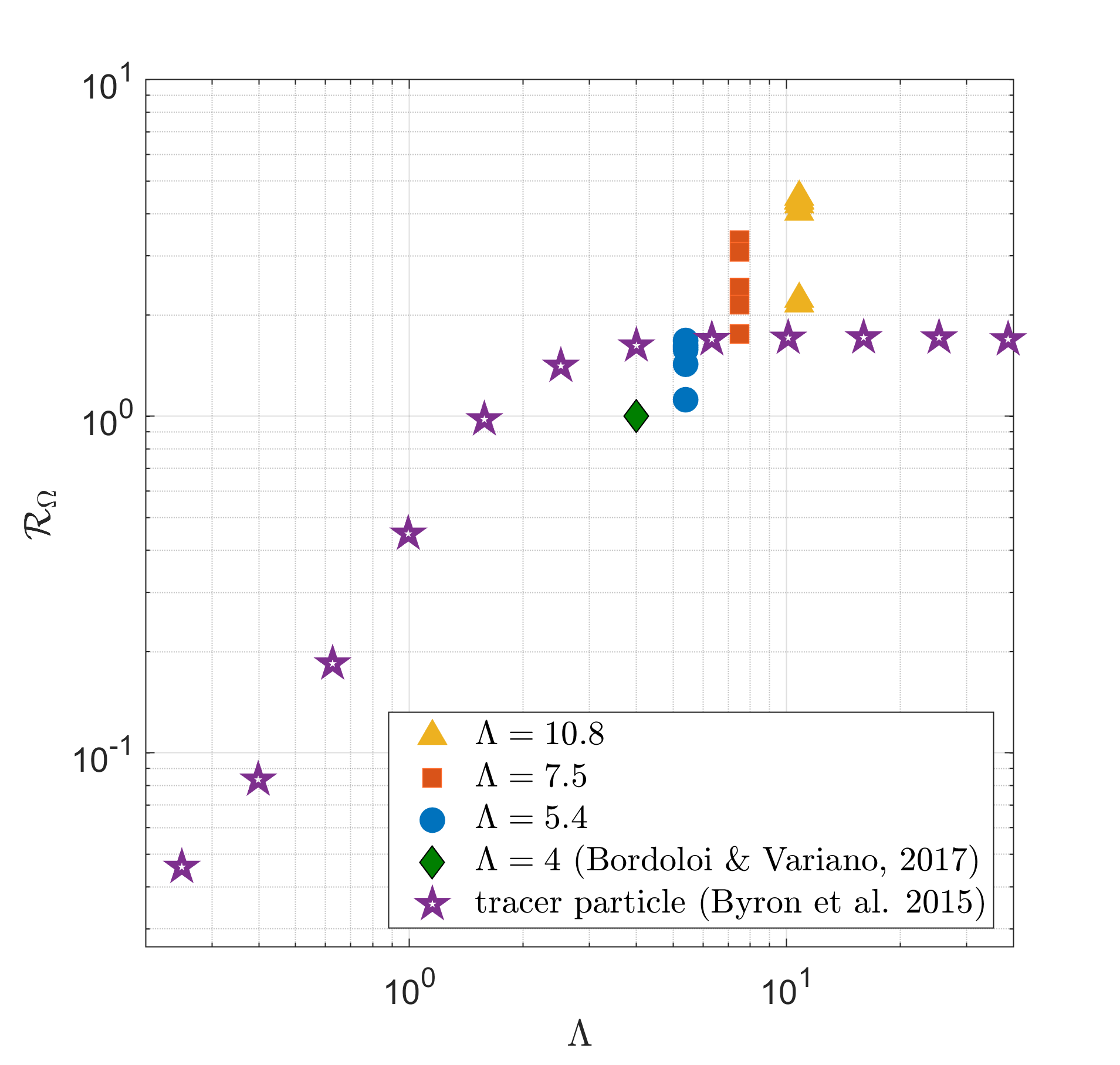}
		\caption{}
		\label{fig:7a}
      \end{subfigure}
	\begin{subfigure}[b]{0.45\textwidth}
		\includegraphics[width=\linewidth]{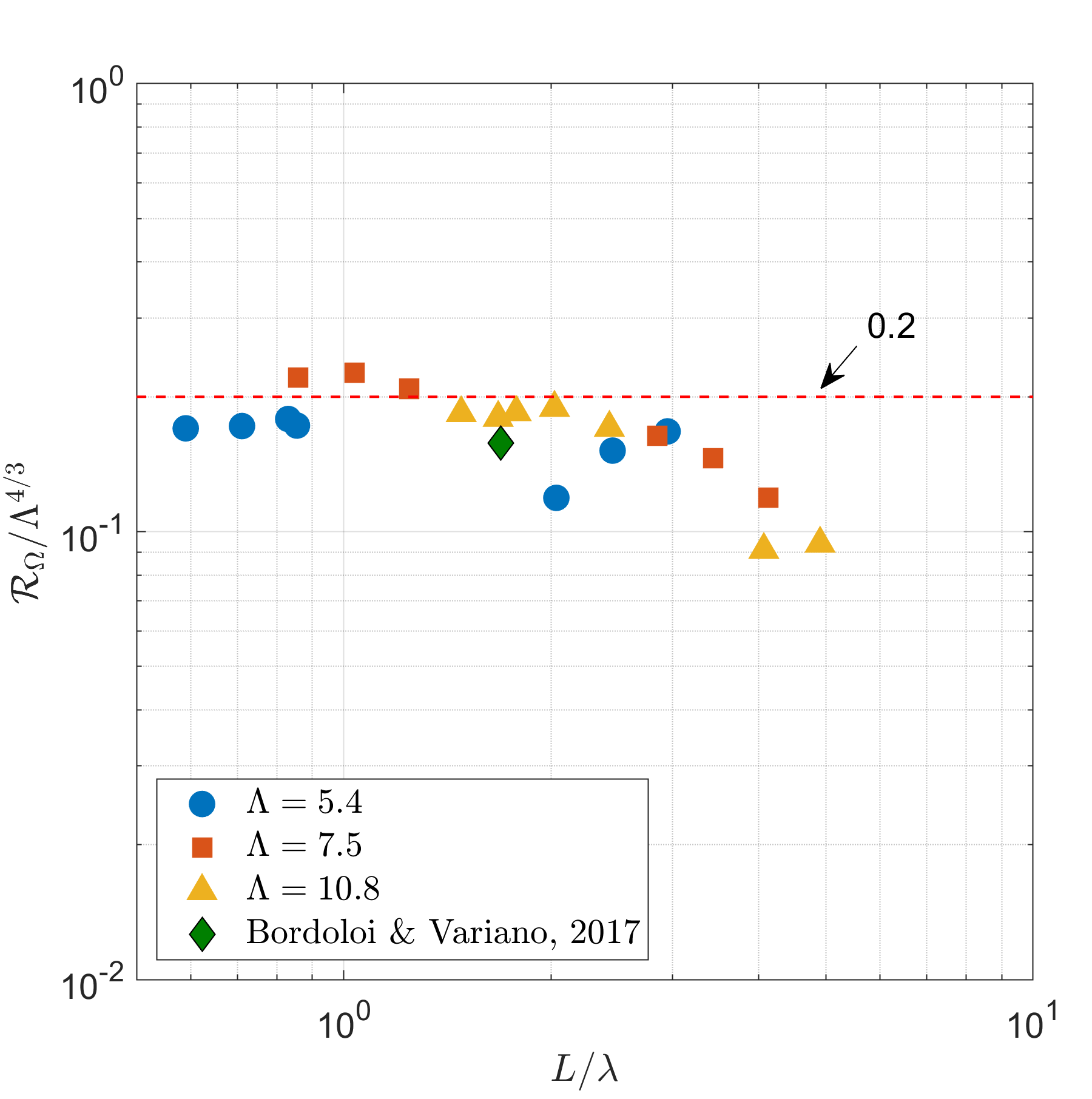}
		\caption{}
		\label{fig:7b}
      \end{subfigure}
		\caption{a) Ratio between the variances of the spinning and the tumbling rate ($\mathcal{R}_{\Omega}=\langle\Omega_S\Omega_S\rangle/\langle\Omega_T\Omega_T\rangle$) against the fiber aspect ratio and b) a compensated ratio ($\mathcal{R}_{\Omega}/\Lambda^{4/3}$) against fiber length $L$ normalized by Taylor length $\lambda$.}
		\label{Fig:TotalRotation1}
	\end{center}
\end{figure} 

Fibers with sub-Kolmogorov length ($L<\eta_K$) act as material lines and preferentially align in the direction of fluid vorticity~\citep{Byron2015}. This leads such fibers to show stronger spinning than tumbling, such that $\mathcal{R}_{\Omega} = \langle \Omega_S\Omega_S\rangle/\langle\Omega_T\Omega_T\rangle >1$ when $\Lambda>1$. FIG~\ref{fig:7a} compares the evolution of $\mathcal{R}_{\Omega}$ between sub-Kolmogorov particles and current measurements as a function of $\Lambda$. For sub-Kolmogorov particles, $\mathcal{R}_{\Omega}$ increases with $\Lambda$ as the shape of a particle changes from disc ($\Lambda<1$) to rod ($\Lambda>1$) and becomes nearly constant at $\mathcal{R}_{\Omega} \sim 1.7$ when $\Lambda \geq 6$. \citet{Bordoloi2017} reported that for cylinders with $\Lambda=4$, $\mathcal{R}_{\Omega}=1$. These data points are shown with our current measurements for large fibers in figure \ref{fig:7a}. The combined result shows that $\mathcal{R}_{\Omega}$ continues to increase from 1 up to 4.5 when $\Lambda$ increases to 10.8.

The evolution of the spinning rate with the fiber diameter discussed above cannot be explained only on the basis of the classical K41 approach. Indeed, within this framework, a structure of size $\ell$ is correlated over the size of order $\ell$. Therefore, the integral over a length greater than $\ell$ vanishes, as $\langle u_\ell \rangle =0$. This is obviously not the case here.   The observed scaling of the spinning rate implies that fibers might be preferentially aligned with elongated structures where transverse increments of velocity are correlated over a longer lengthscale. In turbulent flows, such structures exist typically as the filaments of coherent vorticity as first evidenced by~\citet{Douady1991}. These coherent structures can be very long, up-to the integral length of the flow, but are generally twisted and randomly oriented. The forcing of the spinning is then only possible as long as the fiber length is smaller or of the order of the correlation length of the axial vorticity of these filaments. \citet{Jimenez1998} showed numerically that this correlation length is given by the Taylor lengthscale $\lambda$. Such preferential alignment of elongated particle with coherent vortices has also been recently reported by \citet{Picardo2020} for flexible fibers.

We examine this hypothesis in FIG.~\ref{fig:7b} that shows the evolution of the compensated ratio $\mathcal{R}/ \Lambda^{4/3}$ with respect to fiber length ($L$) normalized by the Taylor lengthscale ($\lambda$). Result shows that $\mathcal{R}_{\Omega}/\Lambda^{4/3}$ is nearly constant and equal to 0.2 up to $L \sim 2\lambda$ after which it continues to decrease.  This supports the argument that the forcing of the spinning is due to coherent structure which correlation length scales with the Taylor scale.

\begin{figure}
	\begin{center}
		\centering
	\begin{subfigure}[b]{0.45\textwidth}
		\includegraphics[width=\linewidth]{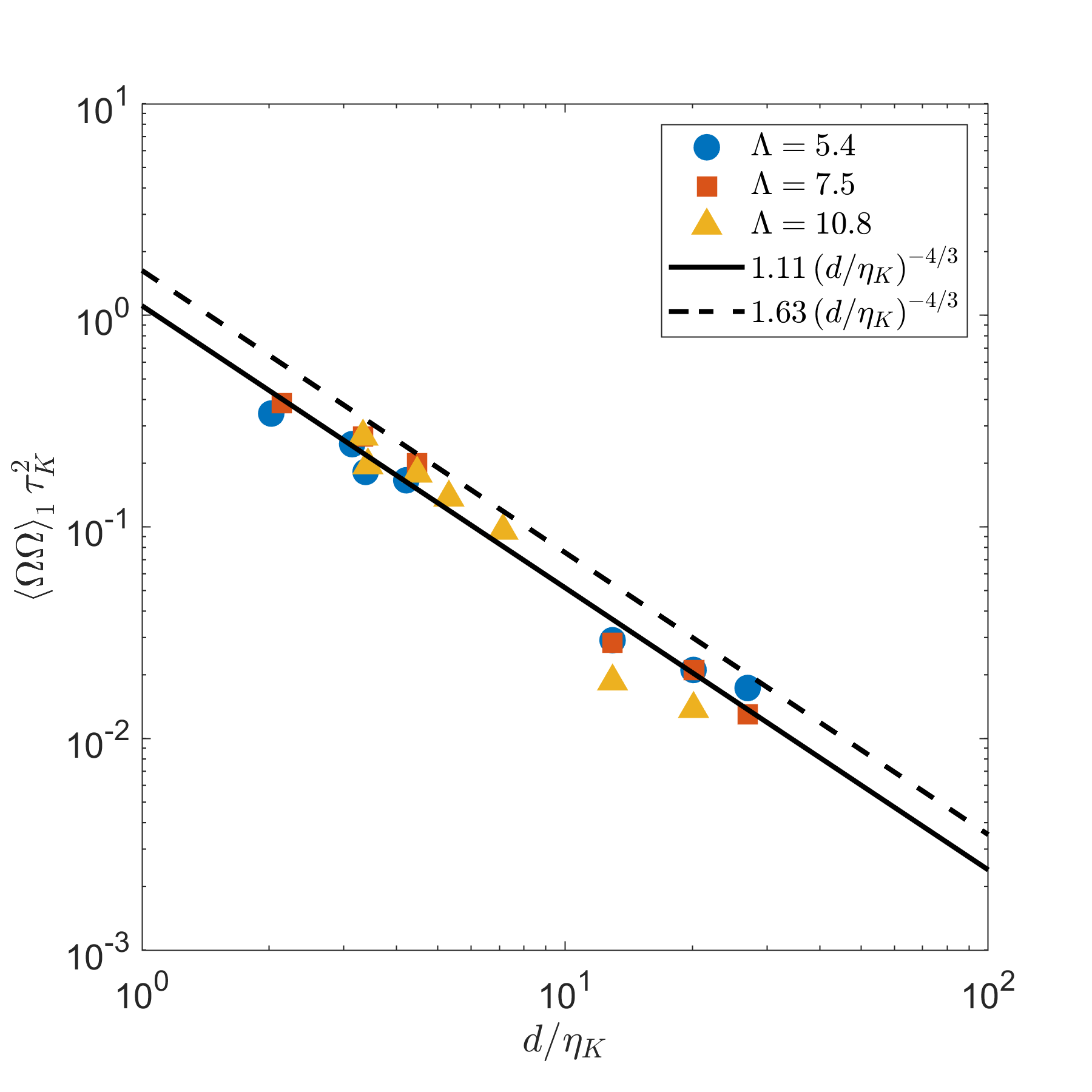}
		\caption{}
		\label{fig:6a}
      \end{subfigure}
	\begin{subfigure}[b]{0.45\textwidth}
		\includegraphics[width=\linewidth]{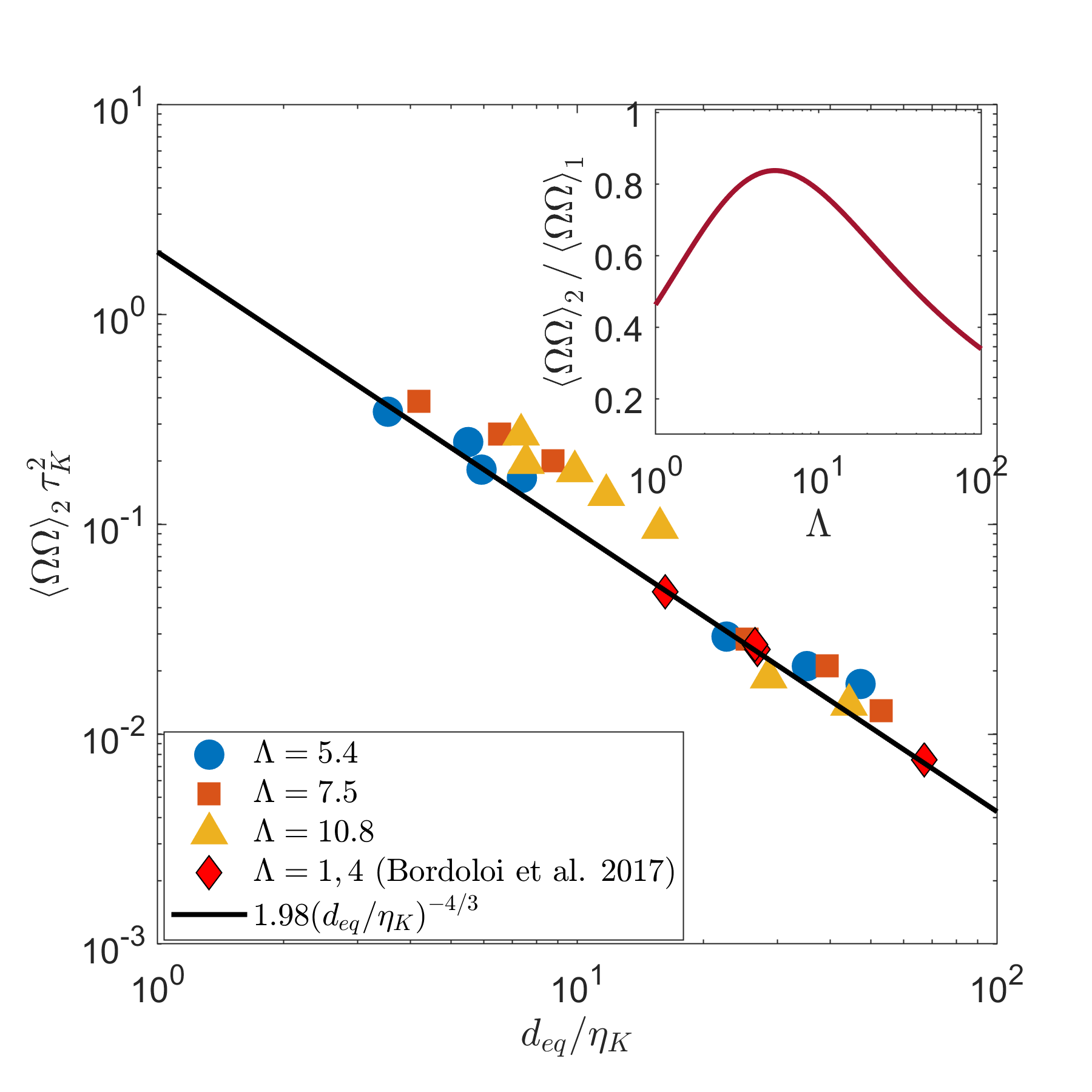}
		\caption{}
		\label{fig:6b}
      \end{subfigure}
		\caption{Dimensionless variance of total rotation rate against dimensionless a) fiber diameter based on equation \ref{eq:totrot1}, b) spherical volume equivalent diameter based on equation \ref{eq:totrot2} and the evolution of the ratio between the two variances as a function of aspect ratio in the inset.}
		\label{Fig:TotalRotation2}
	\end{center}
\end{figure}

FIG.~\ref{fig:6a} shows the normalized variance of total rotation rate, $\langle\Omega\Omega\rangle\tau_K^2 = \left(\langle\Omega_S\Omega_S\rangle + \langle\Omega_T\Omega_T\rangle\right)\tau_K^2$ with respect to the normalized fiber diameter ($d/\eta_K$). Combining the scalings of the variances of spinning and tumbling rates with respect to fiber diameter and length, the variance of total rotation rate can be expressed as,
\begin{align}
    \langle\Omega\Omega\rangle_1\tau_K^2&= \langle\Omega_S\Omega_S\rangle\tau_K^2 + \langle\Omega_T\Omega_T\rangle \tau_K^2\nonumber\\
    &=C_S\left(d/\eta_K\right)^{-4/3}\left(1 + C_T/C_S\Lambda^{-4/3}\right).
    \label{eq:totrot1}
\end{align}
\noindent
Here, $C_T=4.06$ and $C_S=0.77$ are the two constants of proportionality corresponding to the best fit shown on FIG.~\ref{fig:03} and \ref{fig:5}.
Equation~\ref{eq:totrot1} suggests that $\langle\Omega\Omega\rangle$ follows a -4/3 power-law with respect to $d/\eta_K$, and the prefactor depends on the aspect ratio $\Lambda$. For the three aspect ratios considered in this study, the prefactor varies between 1.1 and 1.63 (shown by a solid and a dashed line in FIG.~\ref{fig:6a}).

\citet{Bordoloi2017} found empirically that the evolution of the variance of the total rotation is well described by a power law $\langle \Omega\Omega \rangle \sim \tau_K^{-2}(d_eq/\eta_K)^{-4/3}$, where $d_{eq} \sim d\Lambda^{1/3}$ is the volume equivalent spherical diameter. This relation ca be rewritten as:
\begin{equation}
    \langle\Omega\Omega\rangle_2\tau_K^2 = C\left(d/\eta_K\right)^{-4/3}\Lambda^{-4/9}.
    \label{eq:totrot2}
\end{equation}
FIG.~\ref{fig:6b} shows also a good agreement to the $d_{eq}/\eta_K$ scaling for the aspect ratios ($\Lambda = 5.4, 7.5, 10.8$) considered in these studies and the aspect ratio presented in~\citet{Bordoloi2017} ($\Lambda = 1, 4$). The solid line in this plot shows the power law fit $1.98\left(d_{eq}/\eta_K\right)^{-4/3}$ proposed in \citet{Bordoloi2017}. The equivalence of these two scaling laws is captured by the ratio $\langle\Omega\Omega\rangle_2/\langle\Omega\Omega\rangle_1$ shown in the inset of FIG.~\ref{fig:6b}. For the $\Lambda$ values considered in these studies, the ratio between the two variances is close to 1, such that $\langle\Omega\Omega\rangle_2/\langle\Omega\Omega\rangle_1$ ranges between 0.5-0.84 for $\Lambda = 1-10$. For larger aspect ratio, we expect that the scaling previously proposed by \citet{Bordoloi2017} underestimates the total rotation $\langle \Omega \Omega \rangle$.

	\subsection{Lagrangian time-scales and intermittency}

\begin{figure}
	\begin{center}
		\centering
\begin{subfigure}[b]{0.45\textwidth}
		\includegraphics[width=\linewidth]{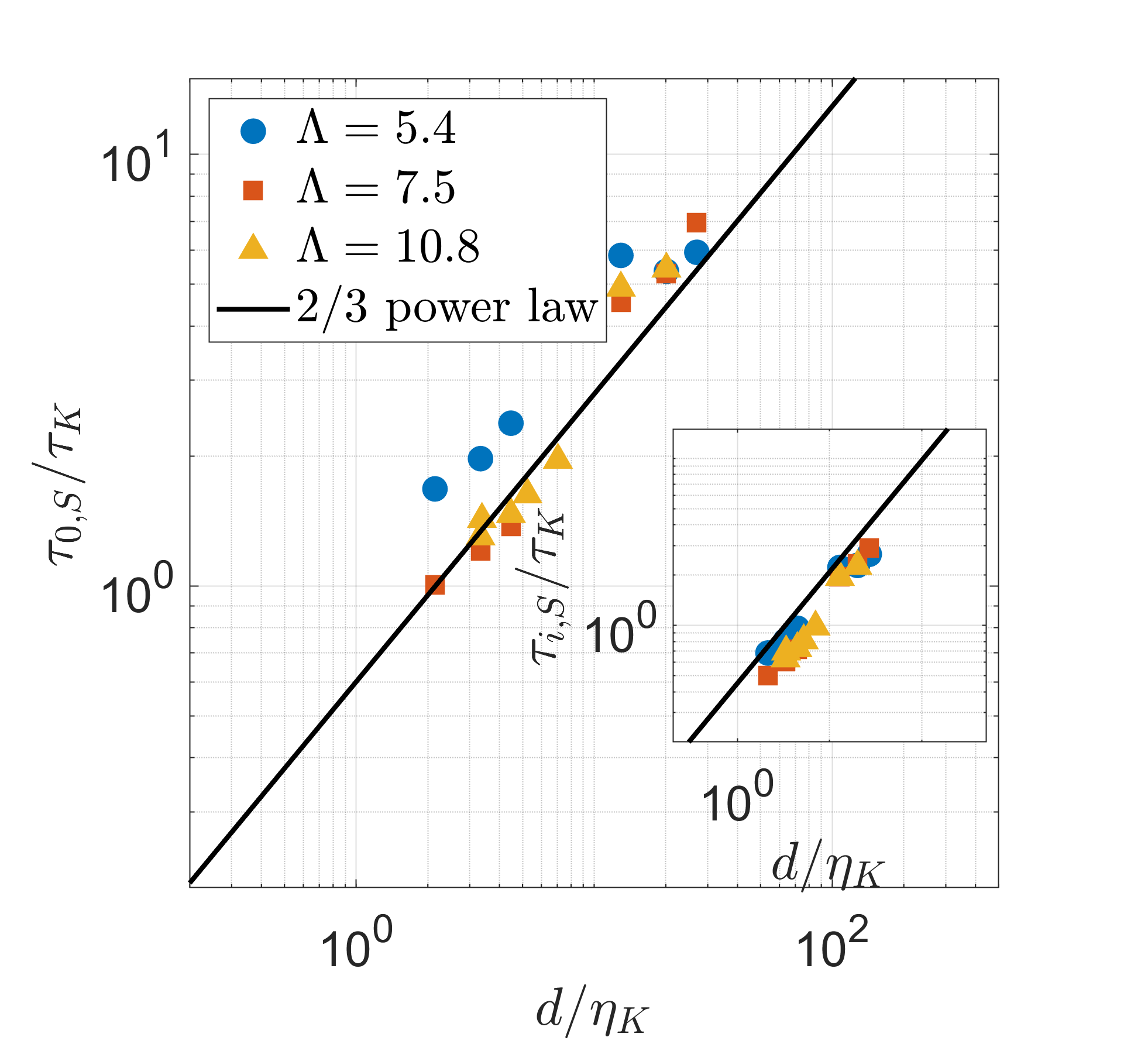}
		\caption{}
		\label{fig:8a}
      \end{subfigure}
	\begin{subfigure}[b]{0.45\textwidth}
		\includegraphics[width=\linewidth]{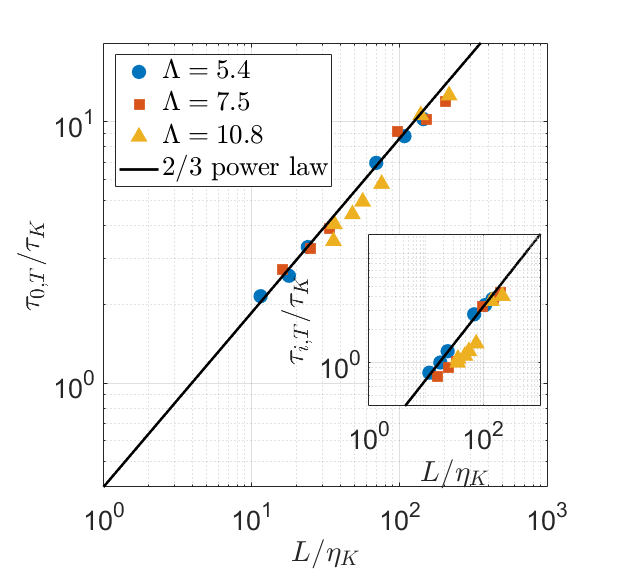}
		\caption{}
		\label{fig:8b}
      \end{subfigure}

		\caption{Evolution of the zero-crossing time for a) tumbling rate as a function of the normalized fiber length ($L/\eta_K$) and b) spinning rate as a function of the normalized fiber diameter ($d/\eta_K$). Each inset shows the evolution of the integral correlation time-scale for the respective component of rotation.}
		\label{Fig:correltime}
	\end{center}
\end{figure}

 Given that the variance of spinning rate of a fiber scales with the fiber diameter ($d/\eta_K$) and not the fiber length ($L/\eta_K$), the goal here is to probe if such scaling also exists for the correlation time  of the spinning rate. Following the method described in \citet{Bordoloi2020}, we compute two time-scales, namely the zero-crossing time ($\tau_0$) and the integral time ($\tau_i$), based on the mean autocorrelation of the spinning and the tumbling rates from approximately 1500 trajectories. The details of this computation can be found in \citet{Bordoloi2020}.

FIG.~\ref{fig:8a} recovers the trend observed in \citet{Bordoloi2020} and shows that, the evolution of the normalized correlation times ($\tau_{0,t}/\tau_K$ and $\tau_{i,t}/\tau_K$) of the tumbling rate with the normalized fiber length ($L/\eta_K$) collapses on a power-law, 
$\tau_{i}/\tau_K \sim \tau_0/\tau_K \sim (L/\eta_K)^{2/3}$. This result reemphasizes that the fiber length ($L/\eta_K$) characterizes not only the variance, but also the Lagrangian time-scale of the tumbling rate, and that the fiber diameter ($d/\eta_K$) has no significant role when $St_T <1$. A similar 2/3-power law scaling is recovered for the normalized correlation times ($\tau_{0,s}/\tau_K, \tau_{i,s}/\tau_K$) of the spinning rate when plotted with respect to the normalized fiber diameter ($d/\eta_K$) (see FIG.~\ref{fig:8b}). We do not observe any systematic deviation from the power-law scaling in the correlation time-scales of the spinning rate for all the tested diameter, unlike the correlation time-scales of tumbling rate in~\citet{Bordoloi2020} which depend on a tumbling Stokes number.  Nonetheless, this result confirms that the normalized fiber diameter is an important length scale of the coherent structures that force spinning.

\begin{figure}
	\begin{center}
		\centering
\begin{subfigure}[b]{0.45\textwidth}
		\includegraphics[width=\linewidth]{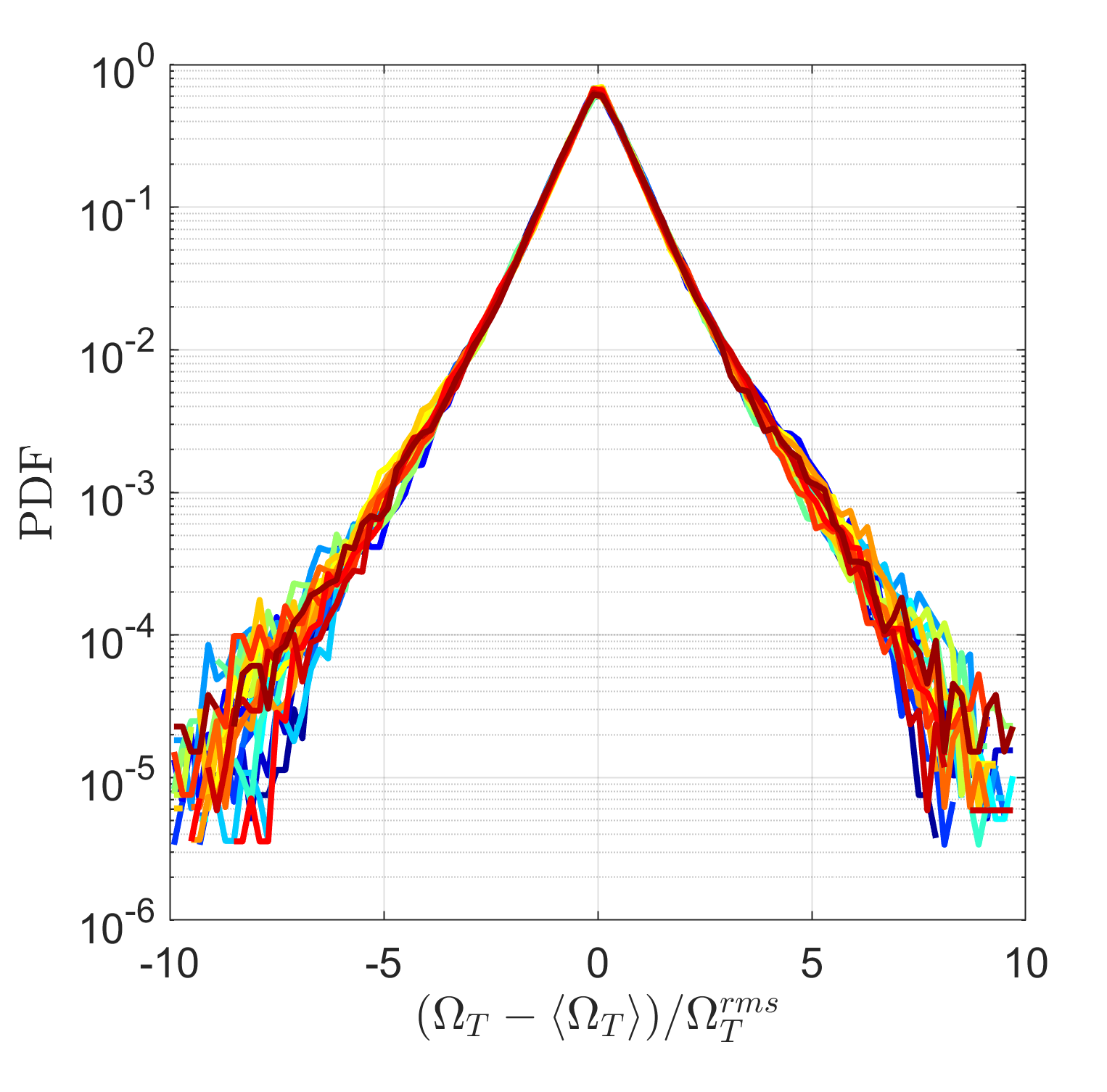}
		\caption{}
		\label{fig:9a}
      \end{subfigure}
	\begin{subfigure}[b]{0.45\textwidth}
		\includegraphics[width=\linewidth]{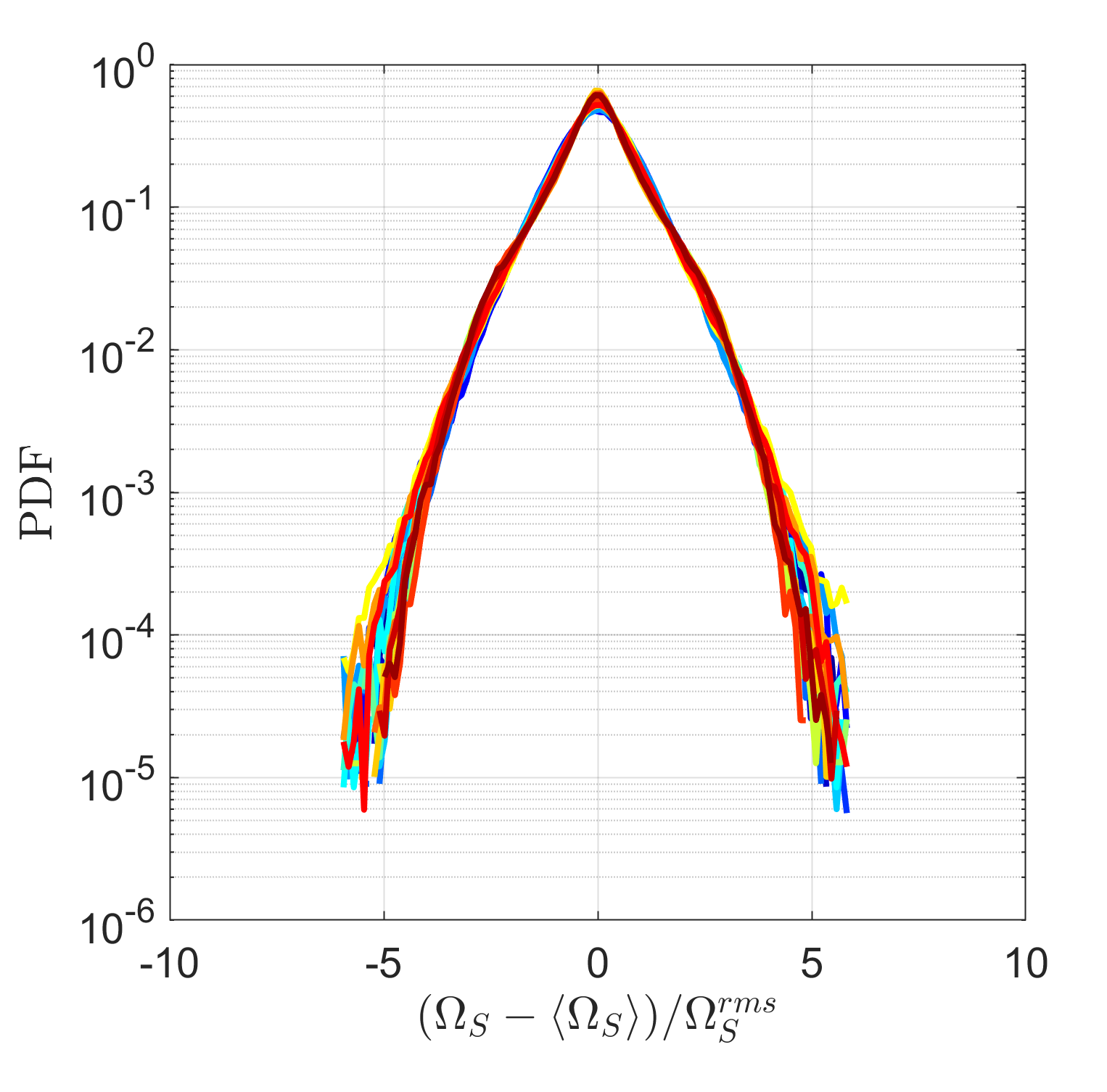}
		\caption{}
		\label{fig:9b}
      \end{subfigure}

		\caption{PDF of mean-subtracted a) tumbling and b) spinning rates normalized by their respective R.M.S. for fibers of various lengths and diameter. The color-scheme varies from blue to red with increasing $L/\eta_K$ and $d/\eta_K$ for tumbling and spinning, respectively.}
		\label{Fig:pdf}
	\end{center}
\end{figure}

FIG.~\ref{fig:9a} and \ref{fig:9b} present the probability density function (PDF) of the tumbling and the spinning rates, respectively. Each distribution is mean-centered and normalized by the R.M.S. of the respective component. The mean spinning and tumbling rates are close to zero and much smaller than the respective variance. The distributions are independent of their Cartesian components and hence, only their vertical component is shown for tumbling. 

Results show that the shape of the PDF for both tumbling and spinning are independent of the fiber size within the errorbar of our measurements. For the tumbling rate, the PDFs are symmetric, and they show exponential decay. This shape is compatible with the one observed by \citet{Parsa2014}, who presented the PDF of the norm of the tumbling rate. By contrast, the PDFs of the spinning rate are relatively wider for small rotation rates, but then a sharp decay appears for spinning rates larger than $\sim 3\Omega_S^{R.M.S}$.

Vortex tubes are associated with turbulence intermittency responsible for broadening the PDF of the velocity increments $\delta_lu$, the level of intermittency, \textit{i.e.} the flatness, increasing with decreasing length scale $\ell$~\citep{Frisch1995}. If a fiber acts as a proxy of turbulent structures, each component of its rotation is induced by the velocity increment at its respective length scale. This suggests that a smaller length-scale ($d$) should result in a broader stretching of tails for the PDF of the spinning rate compared to that for the tumbling rate induced by a larger length-scale ($L$). Consequently, the flatness ($\mathcal{F}_x=\langle (x-\langle x \rangle )^4 \rangle/ (\langle x-\langle x \rangle)^2 \rangle^2$) of the PDF for the spinning rate should be larger than that for the tumbling rate. In our experiments with inertial fibers, We find that the flatness for spinning ($\mathcal{F}_S\sim 4$) is almost half of the flatness for tumbling ($\mathcal{F}_T\sim 8$). 

This surprising trend with $\mathcal{F}_S < \mathcal{F}_T$ may possibly be due to a filtering effect that suppresses the strongest events responsible for the the tails in the PDF of spinning. Two possible filtering mechanisms are:  a) fiber inertia, and b) decorrelation of forcing along the fiber length. In FIG.~   \ref{fig:9b}, we do not observe any difference between the PDFs for the different $d/\eta_K$. Moreover, we do not observe any effect of inertia on the Lagrangian time-scales of spinning rate (see FIG.~\ref{fig:8b}). This suggests that the fiber inertia is not responsible for the above filtering effect. We relate the damping of the tails in FIG.~\ref{fig:9b} to the second mechanism, that is the filtering due to the sum of all the local contributions to the torque along the fiber length (see equation \ref{eq:04}). One implication of this hypothesis is that, the strongest events of velocity increment in turbulence at scale $d$ are not correlated over length scales $\ell \gg d$. This is compatible with the classical view of turbulence velocity field where the size of the vortex tube decreases with increasing the threshold of the amplitude of enstrophy~\citep{Moisy2004}. To test this idea, experiments that simultaneously measure the dynamics of a fiber and the flow around it should be performed in the future.

	\subsection{Rotation rate and fiber inertia}

Finally, we quantify here the role of inertia on the rotational dynamics of fibers. The general equation of conservation of the angular momentum is:
\begin{equation}
	\frac{\partial I\Omega}{\partial t} + \Omega \times I\Omega = \Gamma
	\label{eq:ConservationKineticMomentum}
\end{equation}
To fully characterize the importance of fiber inertia (terms on the left hand side of equation~\ref{eq:ConservationKineticMomentum}), it is necessary to measure the total torque ($\Gamma$) experienced by the fiber, {which also requires the measurement of the flow{-field}}. Direct measurement of $\Gamma$ is beyond the scope of our current study. However, it is possible to quantify the influence of fiber inertia based on the Stokes number as proposed in~\citet{Bounoua2018}. For the tumbling rate, we showed that the fiber inertia is negligible as long as the tumbling Stokes number is small ($St_T<1$). In a similar vein, we define a spinning Stokes number that compares the time scale of the forcing at the scale of the fiber diameter (\emph{i.e.} $\tau_d\sim d/u_d$) and the relaxation time of the spinning rate $\tau_S\sim I_S/\mu Ld^2$. The relaxation time of the spinning rate is defined by the balance of the inertial term $\partial_t I_S \Omega_S$ and the dissipation term in the conservation of the angular momentum equation which scales as $\mu d^2 L \Omega_S$ considering a viscous torque. This provides a definition of the spinning Stokes number as,
\begin{equation}
	St_S=\frac{\tau_S}{\tau_d} \sim \frac{\pi}{32} \frac{\rho_p}{\rho_f} \left( \frac{d}{\eta_K} \right)^{4/3}.
\label{eq:9}
\end{equation}
The spinning Stokes number for the current dataset varies within the range $St_S\in [0.25 ; 8.2]$. The variance of the spinning rate likely decreases above a critical spinning Stokes number ($St_S$) due to fiber inertia analogous to what was shown in~\citet{Bounoua2018} for tumbling. This threshold depends on the prefactor used to compute the relaxation time ($\tau_S$), which we take equal to 1 for our analysis.
For the present set of experiments, if a filtering due to the inertia exists it is weak as the evolutions of both the variance, the correlation time and the pdf do not seem to depend on the Stokes number.

In~\citet{Bounoua2018}, such inertial filtering on the tumbling rate was captured by a low pass filter whose transfer function depends solely on the tumbling Stokes number $St_T$.  This approach however is not as straightforward for the spinning rate as we claim that the forcing of the spinning is due to {a) the increment of velocity at the scale of the diameter of the fiber, and b) the ratio $L/\lambda$ because of} the correlation of the axial vorticity of the vortex tube along the fiber length (see section~\ref{sc:3C}). To probe the influence of fiber inertia on spinning by avoiding the influence of such correlation requires to have a separation of scales between the fiber lengthscales and the flow lengthscales such that $\eta_K \lesssim d < L \ll \lambda$, which is not possible in our experimental setup.

The second inertial term is the Coriolis term $\Omega\times I\Omega$. In earlier related studies \citep{Parsa2014, Voth2017, Bounoua2018}, this nonlinear term was neglected for simplicity, and the spinning rate was assumed to be low. However, our results show that the spinning rate is larger than the tumbling rate, raising questions about the validity of this assumption. In the fiber frame, the components of the rotation vector $\Omega$ are $\Omega_{T,1}$, $\Omega_{T,2}$ and $\Omega_{s}$. Hence the components of the Coriolis term for an axisymmetric fiber are $(I_T-I_S) \Omega_S\Omega_{T,2}$, $(I_S-I_T)\Omega_S\Omega_{T,1}$ and 0. Here, $I_T$ and $I_S$ are the moment of inertia of the fiber along the tumbling and the spinning axes, respectively. The last Coriolis term in the direction of spinning axis vanishes due to the axisymmetry condition, i.e. $I_{T,1} = I_{T,2} = I_{T}$. For fibers with $I_S/I_T\sim (d/L)^{2} \ll 1$, the two non-zero Coriolis terms in the direction of tumbling can be approximated as $I_T\Omega_S\Omega_{T,2}$ and $-I_T\Omega_S\Omega_{T,1}$.  These two Coriolis terms are negligible compared to the temporal term $\partial_T I_T\Omega_T$ and the viscous torque as long as:
\begin{equation}
	\Omega_S\tau_T \ll 1.
\label{eq:10}
\end{equation}
Here $\tau_T$ is the relaxation time of tumbling, defined by the balance of the viscous relaxation term $\int \mu L\Omega_T \times s {\rm d} s$ and the temporal term, and scales as $\tau_T\sim I_T/\mu L^3$~\citep{Bounoua2018}. Invoking the scaling for the spinning rate as $\Omega_S\sim u_d/d \sim (d/\eta_K)^{-2/3}\tau_K^{-1}$ into equation \ref{eq:10}, we can show that the Coriolis term is negligible as long as the spinning Stokes number is small $St_S\ll1$. Contrary to the tumbling Stokes number $St_T$, the spinning Stokes number $St_S$ does not depend on the aspect ratio of the fiber and can be significantly large even for slender body if $d>>\eta$  (see equation \ref{eq:9}). Therefore, for fibers with $d>\eta_K$ and $\Lambda\gg1$, although the temporal inertial term $\partial_T I_T\Omega_T$ can be ignored in equation \ref{eq:10}, the Coriolis term cannot always be neglected and couple the tumbling and the spinning of the fiber.

\section{Conclusion}
\label{sec04}

We experimentally resolve both components of rotation (spinning and tumbling) of inertial fibers ($L \gg \eta_K$; $d>\eta_K$) in homogeneous isotropic turbulence. Our measurements show that fibers tend to spin more than to tumble. We show that the variance of the spinning rate follows a power law scaling with respect to the fiber diameter, such that $\langle\Omega_S\Omega_S\rangle\tau_K^{2} \sim (d/\eta_K)^{-4/3}$.  This contradicts the classical view based on K41 theory where the spinning rate of an inertial fiber is considered negligible compared to the tumbling rate. This scaling implies that fibers are preferentially trapped within elongated coherent structures where the transverse increments of velocity are correlated over lengths of the order of Taylor scale of turbulence. We show the importance of the fiber aspect ratio ($\Lambda = L/d$) via a rescaled ratio ($\langle\Omega_S\Omega_S\rangle/\Lambda^{4/3}\langle\Omega_T\Omega_T\rangle$) between the variances of the spinning and the tumbling rates.  For a fiber shorter than a few Taylor scale ($L \lesssim 2\lambda$), this ratio is nearly constant, but decreases rapidly with increasing fiber length. In the future, it would be useful to extend this study to oblate anisotropic particles, such as discs, to examine if the major axes of all anisotropic inertial particles tend to align with the vorticity, similar to sub-Kolmogorov scale particles \citep{Voth2017,Chevillard2013}. Besides, it would also be interesting to study such phenomenon for flexible fibers that can conform to the topology of a vortex tube, in the vein of~\citet{Picardo2020}.

In addition, we compute the Lagrangian time scales of spinning and tumbling by analyzing the autocorrelation of the respective components. Both time scales follow the scaling $\tau_{S/T} \sim (l/\eta_K)^{-2/3}$, where $l = L$ and $d$ for tumbling and spinning, respectively. We do not observe any obvious deviation from this scaling for spinning even for high Stokes number $St_S\sim 8$. Further, the PDF of spinning rate shows suppression of extreme events beyond $3\Omega_S^{R.M.S.}$ leading to smaller flatness factor ($\mathcal{F}_S<\mathcal{F}_T$). We hypothesize that this result is due to the length-wise decorrelation of local forcing responsible of the stronger spinning events. 

 The measurement of the spinning rate also allows us to estimate the importance of the Coriolis force which was generally assumed negligible in earlier studies of fiber rotation in turbulence. We show that this assumption should hold only when the spinning Stokes number $St_S$ is small enough.

 \clearpage
 \bibliographystyle{apsrev4-1} 
\bibliography{PRF_2020}

\end{document}